%% file: tbd.tex
\documentclass[10pt,journal,compsoc]{IEEEtran}

\ifCLASSOPTIONcompsoc
  \usepackage[nocompress]{cite}
\else
  \usepackage{cite}
\fi

\ifCLASSINFOpdf
\else
\fi

\usepackage{amssymb}
\usepackage{amsfonts}
\usepackage{amsmath}
\usepackage{amsthm}
\usepackage{stmaryrd}
\usepackage{multirow}
\usepackage{algorithm}
\usepackage[noend]{algpseudocode} 
\usepackage{balance}
\usepackage{graphicx}
\usepackage{enumerate}
\usepackage{array}
\usepackage{url}
\usepackage{color}

\usepackage{paralist}

\newtheorem{definition}{Definition}

\theoremstyle{definition}

\newcommand{\hide}[1]{}
\newcolumntype{x}[1]{>{\centering\arraybackslash}p{#1}}

\begin{document}
%
\title{Benchmarking Blocking Algorithms \\ for Web Entities}
%
%
%
%

\author{Vasilis~Efthymiou, Kostas~Stefanidis, Vassilis~Christophides
\IEEEcompsocitemizethanks{
\IEEEcompsocthanksitem V. Efthymiou is with the University of Crete and ICS-FORTH, Greece.\protect\\
E-mail: \{vefthym,kstef\}@ics.forth.gr
\IEEEcompsocthanksitem K. Stefanidis is with the University of Tampere, Finland (work done while at ICS-FORTH).\protect\\
E-mail: ksts.stefanidis@gmail.com
\IEEEcompsocthanksitem V. Christophides is with the University of Crete, Greece, and INRIA, Paris-Rocquencourt, France.\protect\\
E-mail: Vassilis.Christophides@inria.fr}
\thanks{Manuscript received Month dd, 2016; revised Month dd, 2016.}}%


\markboth{IEEE Transactions on Big Data,~Vol.~x, No.~x, Month~2016}%
{Efthymiou \MakeLowercase{\textit{et al.}}: Benchmarking Blocking Algorithms for Web Entities}
%



\IEEEtitleabstractindextext{%
\begin{abstract}
An increasing number of entities are described by interlinked data rather than documents on the Web. Entity Resolution (ER) aims to identify descriptions of the same real-world entity within one or across knowledge bases in the Web of data. To reduce the required number of pairwise comparisons among descriptions, ER methods typically perform a pre-processing step, called \emph{blocking}, which places similar entity descriptions into blocks and thus only compare descriptions within the same block. We experimentally evaluate several blocking methods proposed for the Web of data using real datasets, whose characteristics significantly impact their effectiveness and efficiency. The proposed experimental evaluation framework allows us to better understand the characteristics of the missed matching entity descriptions and contrast them with ground truth obtained from different kinds of relatedness links.
\end{abstract}

}

\maketitle

\IEEEdisplaynontitleabstractindextext

%
\IEEEpeerreviewmaketitle

\input{introduction}
\input{preliminaries}
\input{implementation}
\input{experiments_setup}
\input{experiments_evaluation}
\input{related_work}

\input{summary}

\vspace{0.05in}
\noindent\textbf{Acknowledgements:} This work was partially supported by the EU FP7 SemData (\#612551) project.




\ifCLASSOPTIONcaptionsoff
  \newpage
\fi




\bibliographystyle{IEEEtran}
\bibliography{jws_ref}

%

\begin{IEEEbiographynophoto}
{Vasilis Efthymiou}
is a PhD candidate at the University of Crete, Greece, and a research assistant at ICS-FORTH, Greece. The topic of his PhD is entity resolution in the Web of data. He got his MSc and BSc degrees from the same university in 2012 and 2010, respectively. He has received undergraduate and postgraduate scholarships from FORTH, working in the areas of Semantic Web 
and Ambient Intelligence.
\end{IEEEbiographynophoto}


\begin{IEEEbiographynophoto}
{Kostas Stefanidis} is an Associate Professor at the University of Tampere, Finland. 
He got his PhD in personalized data management from the University of Ioannina, Greece, in 2009. 
His research interests lie in the intersection of databases, Web and information retrieval. 
Kostas 
co-authored more than 35 papers in peer-reviewed conferences and journals, including ACM SIGMOD, IEEE ICDE and ACM TODS. 
\end{IEEEbiographynophoto}



\begin{IEEEbiographynophoto}
{Vassilis Christophides} is a Professor at the University of Crete, Greece. He has been recently appointed to an advanced research position at INRIA Paris - Rocquencourt. Previously, he worked as Distinguished Scientist at Technicolor, R\&I Center in Paris. He studied Electrical Engineering at the National Technical University of Athens, Greece, 1988, he received his DEA in computer science from the University PARIS VI, 1992, and his Ph.D. from the Conservatoire National des Arts et Metiers of Paris, 1996. His main research interests include Databases and Web Information Systems, as well as Big Data Processing and Analysis. He has published over 130 articles in high-quality international conferences, journals and workshops. He received the 2004 SIGMOD Test of Time Award and the ISWC Best Paper Award in 2003 and 2007. 
\end{IEEEbiographynophoto}



\end{document}

%% file: introduction.tex
\IEEEraisesectionheading{\section{Introduction}\label{sec:introduction}}

%
%
%
%

\IEEEPARstart{O}{ver} the past decade, numerous {\em knowledge bases} (KBs) have been built to power large-scale knowledge sharing, but also an entity-centric Web search, mixing both structured data and text querying. 
These KBs offer comprehensive, machine-readable descriptions of a large variety of real-world entities (e.g., persons, places) published on the Web as {\em Linked Data} (LD).  
Traditionally, KBs are manually crafted by a dedicated team of knowledge engineers, such as the pioneering projects Wordnet and Cyc. Today, more and more KBs are built from existing Web content using information extraction tools. 
Such an automated approach offers an unprecedented opportunity to scale-up KBs construction and leverage existing knowledge published in HTML documents.

Although KBs (e.g., 
DBpedia, 
Freebase
) 
may be derived from the same data source (e.g., Wikipedia), they may provide multiple descriptions of the same entities. This is mainly due to the different information extraction tools and curation policies 
employed by KBs, resulting to complementary and sometimes conflicting descriptions. \emph{Entity resolution} (ER) aims to identify descriptions that refer to the same entity within or across KBs~\cite{DBLP:series/synthesis/2015Christophides,DBLP:series/synthesis/2015Dong}. 
ER is essential in order to improve {\em interlinking} in the Web of data, even by third-parties.
In particular: 
\begin{itemize}
\item The size of the Linking Open Data (LOD) cloud\footnote{\url{http://lod-cloud.net}}, in which nodes are KBs (aka RDF datasets) and edges are links crossing KBs, has roughly doubled between 2011 and 2014~\cite{DBLP:conf/semweb/SchmachtenbergBP14}, while data interlinking dropped by 30\%. In general, the majority of the KBs are sparsely linked, while their popularity in links is heavily skewed\footnote{\url{http://linkeddata.few.vu.nl/wod_analysis}}. Sparsely interlinked KBs appear in the periphery of the LOD cloud (e.g., Open Food Facts, Bio2RDF), while heavily interlinked ones lie at the center (e.g., DBpedia, GeoNames, FOAF). Encyclopaedic KBs, such as DBpedia, or widely used georeferencing KBs, such as GeoNames, are interlinked with the largest number of KBs both from the LOD center and the periphery.
\item The descriptions contained in these KBs present a high degree of {\em semantic} and {\em structural} diversity, even for the same entity types. The former is due to the frequent creation of new names for entities that have been described in another KB, 
as well as the simultaneous annotation of descriptions with semantic types not necessarily originating from the same vocabulary. The latter is due to the diverse sets of properties used to describe entities both in terms of types and number of occurrences, even within a KB.
\end{itemize}

The \emph{scale}, \emph{diversity} and \emph{graph structuring} of entity descriptions in the Web of data challenge the way two descriptions can be effectively compared in order to efficiently decide whether they are referring to the same real-world entity. This clearly requires an understanding of the relationships among \emph{somehow similar} entity descriptions that goes beyond duplicate detection without always being able to merge related descriptions in a KB and thus improve its quality. Furthermore, the {\em very large volume} of entity collections that we need to resolve in the Web of data is prohibitive when examining pairwise all descriptions.


In this context of big Web data, \textit{blocking} is typically used as a pre-processing step for ER to reduce the number of unnecessary comparisons, i.e., comparisons between descriptions that do not match. 
After blocking, each description can be compared only to others placed within the same block. 
The desiderata of blocking are to place (i) similar descriptions in the same block (\emph{effectiveness}), and (ii) dissimilar descriptions in different blocks (\emph{efficiency}). 
However, efficiency dictates skipping many comparisons, possibly leading to many missing matches, which in turn implies low effectiveness. 
Thus, the main objective of blocking is to achieve a trade-off between the number of comparisons suggested and the number of missed matches. 


Most of the blocking algorithms proposed in the literature (for a survey, refer to~\cite{DBLP:books/daglib/0030287}) assume both the availability and knowledge of the schema of the input data, i.e., they refer to relational databases. To support a Web-scale resolution of heterogeneous and loosely structured entities across domains, recent blocking algorithms (e.g., \cite{DBLP:conf/wsdm/PapadakisINPN12,DBLP:journals/tkde/PapadakisIPNN13}) disregard strong assumptions about knowledge of the schema of data and rely on a minimal number of assumptions about how entities match (e.g., when they feature a common token in their description or URI) within or across sources. However, these algorithms have not yet been experimentally evaluated with LOD datasets exhibiting different characteristics in terms of the underlying number of entity types and size of entity descriptions (in terms of property-value pairs), as well as their structural (i.e., property vocabularies) and semantic (i.e., common property values and URLs) overlap.



In summary, in this paper:
\begin{compactitem}
\item We design a large-scale evaluation on a cluster of 15 machines using real data. 
To capture the differences in the heterogeneity and overlap of entity descriptions, we distinguish between data originating from sources in the {\em center} (i.e., heavily interlinked) and the {\em periphery} (i.e., sparsely interlinked) of the LOD cloud.
\item We empirically study the behavior of blocking algorithms for datasets exhibiting different semantic and structural characteristics. We are interested in quantifying the factors that make blocking algorithms take different decisions on whether two descriptions from real LOD sources 
potentially match or not.
\item We investigate typical cases of missed matches of existing blocking algorithms and examine alternative ways for them to be retrieved. We finally present the results of blocking, when other kinds of links, different to {\em owl:sameAs}, are used as a ground truth.
\end{compactitem}

The rest of the paper is organized as follows\footnote{A preliminary abridged version of this paper appeared in~\cite{DBLP:conf/bigdataconf/EfthymiouSC15}.}:
Section~\ref{sec:preliminaries} describes the problem and existing solutions. 
Section~\ref{sec:implementation} presents our implementation of blocking algorithms in MapReduce. Sections~\ref{sec:setup} and~\ref{sec:experiments} analyze the setup and the evaluation results of our experiments, respectively. 
Section~\ref{sec:related_work} overviews works related to ER 
and, finally, Section~\ref{sec:summary} summarizes the paper.

%% file: preliminaries.tex
\section{Blocking Algorithms}
\label{sec:preliminaries}
We consider that an \textit{entity description} is expressed as a set of attribute-value pairs. 
Then, \textit{entity resolution} is the problem of identifying descriptions of the same entity (called \textit{matches}). 
In general, ER can be distinguished between {\em pairwise} and {\em collective}. 
\emph{Pairwise ER} (e.g.,~\cite{DBLP:journals/vldb/BenjellounGMSWW09}) compares two descriptions at a time, depending only on the data contained in these descriptions. 
\emph{Collective ER} (e.g.,~\cite{DBLP:journals/tkdd/BhattacharyaG07}) compares a set of related  descriptions, heavily relying on similarity evidence provided by neighboring descriptions. 

Given as input of ER the descriptions of Figure~\ref{fig:descriptions}, 
$\mathcal{E} = \{e_1, e_2, e_3, e_4, e_5, e_6, e_7\}$, a possible output $P = \{\{e_1, e_6\},$ $\{e_2, e_5\}, \{e_3\}, \{e_4\}, \{e_7\}\}$ indicates that the descriptions $e_1$ and $e_6$ refer to the same real-world object, namely Eiffel Tower, $e_2$ and $e_5$ both represent another object, the Statue of Liberty, and $e_3$, $e_4$ and $e_7$ represent by themselves the entities Auguste Bartholdi, Joan Tower and Bartholdi Fountain, respectively.  
Such a collection is called \textit{dirty}, since it contains duplicates, and the corresponding task is called \textit{dirty ER}, while \textit{clean-clean ER} is a special case of (dirty) ER~\cite{DBLP:conf/edbt/KimL10,DBLP:journals/tkde/PapadakisIPNN13}; $\mathcal{E}$ consists of two clean, i.e., duplicate-free, but possibly overlapping collections, and ER targets at identifying their common descriptions\footnote{\textit{Dirty-dirty} and \textit{clean-dirty ER} can be seen as equivalent to \textit{dirty ER}.  Other names include \textit{record-linkage} (for linking clean KBs) and \textit{deduplication} (for merging duplicates in a dirty KB).}. In this work, we focus on both cases.

Given $\mathcal{E}$, we define a blocking collection as a set of blocks containing the descriptions in $\mathcal{E}$. 
\begin{definition}[Blocking collection]\label{def:blocking}
Let $\mathcal{E}$ be a set of entity descriptions. A blocking collection is a set of blocks $B = \{b_1, \ldots,$ $b_m\}$, such that, 
$\bigcup\limits_{b_i \in B} b_i=\mathcal{E}$.
\end{definition}

\begin{figure}
\scriptsize 
\begin{tabular}{|p{8.4cm}|}
\hline
$e_1$ = \{(about, Eiffel Tower), (architect, Sauvestre), (year, 1889), (located, Paris)\} \\ \hline
$e_2$ = \{(about, Statue of Liberty), (architect, Bartholdi Eiffel), (year, 1886), (located, NY)\} \\ \hline
$e_3$ = \{(about, Auguste Bartholdi), (born, 1834), (work, Paris)\} \\ \hline
$e_4$ = \{(about, Joan Tower), (born, 1938)\} \\ \hline
$e_5$ = \{(work, Lady Liberty), (artist, Bartholdi), (location, NY)\} \\ \hline
$e_6$ = \{(work, Eiffel Tower), (year-constructed, 1889), (location, Paris)\} \\ \hline
$e_7$ = \{(work, Bartholdi Fountain), (year-constructed, 1876), (location, Washington)\} \\ \hline
\end{tabular}
\caption{A set of entity descriptions.}
\label{fig:descriptions}
\vspace{-15pt}
\end{figure}

In general, blocking can be used both for pairwise and collective ER, to reduce the number of required comparisons. Specifically, {\em token 
blocking}~\cite{DBLP:journals/tkde/PapadakisIPNN13} relies on the minimal assumption that matching descriptions should at least share a common token. Each distinct token $t$ in the values of a description, defines a new block $b_t$, essentially building an inverted index of  descriptions. 
Two descriptions are placed in the same block, if they share a token in their values. 

\begin{figure} [t]
\begin{center}
\includegraphics[scale=0.42]{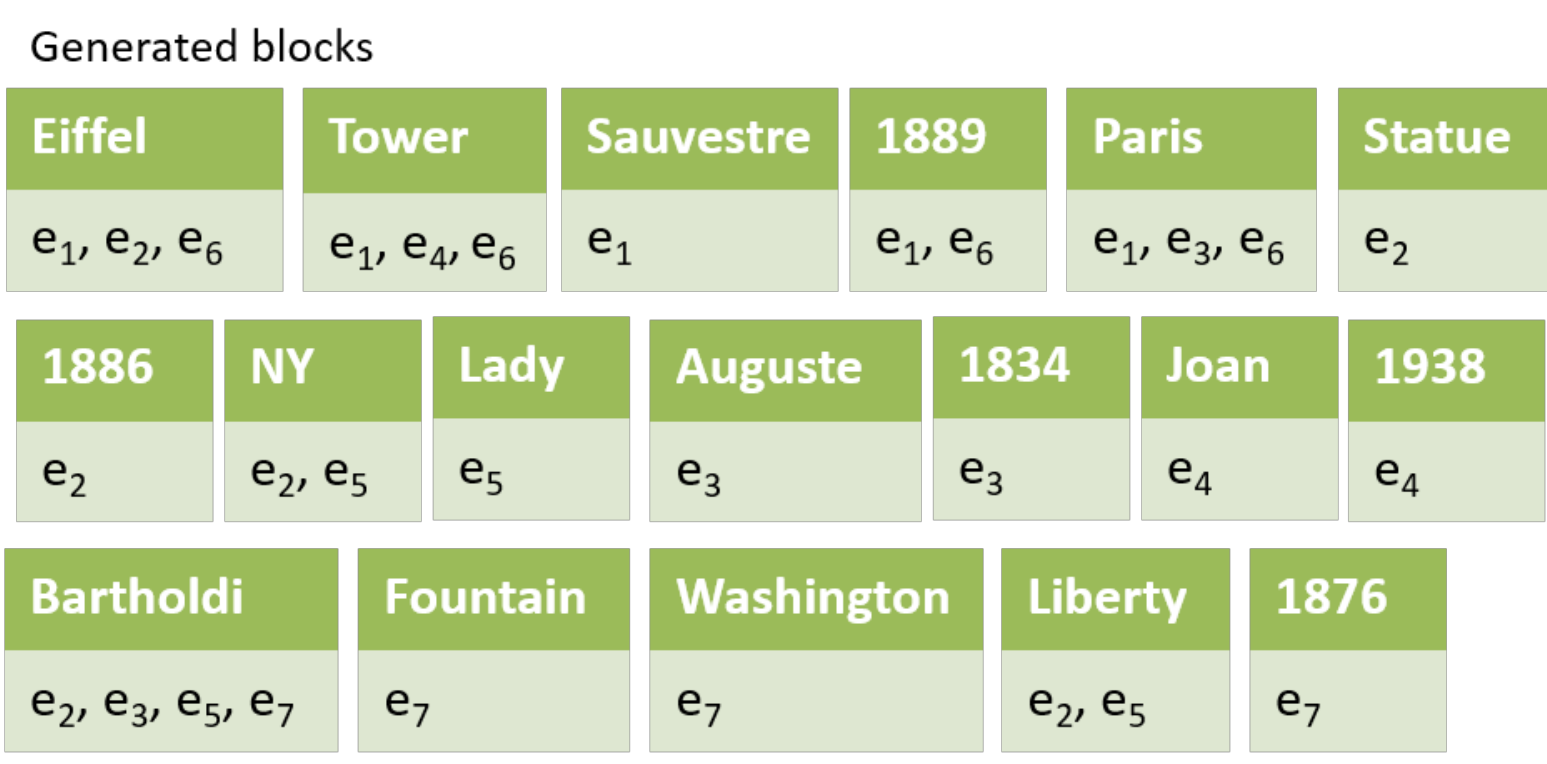} 
\vspace{-11pt}
\caption{Token blocking example. Descriptions having a common token are placed in a common block.}
\label{fig:token_ex}
\vspace{-15pt}
\end{center}
\end{figure}

Given the entity collection of Figure \ref{fig:descriptions}, Figure \ref{fig:token_ex} shows the blocks generated by token blocking. 
In the generated blocks, we save the comparisons ($e_1, e_5$), ($e_1, e_7$), ($e_2, e_4$), ($e_3, e_4$), ($e_4, e_5$), ($e_5, e_6$) and ($e_6, e_7$),
and we successfully place the matches ($e_1, e_6$) and ($e_2, e_5$) in common blocks. Still, pairs, such as ($e_1, e_2$), ($e_1, e_3$), and ($e_3, e_6$),  lead to unnecessary comparisons. Note also that the pair ($e_1, e_6$) is contained in 4 different blocks, which leads to redundant comparisons.

Next, we present two extensions: attribute clustering blocking, in which candidate matches should at least share a common token for similar attributes known globally, and prefix-infix(-suffix) blocking, in which candidate matches should additionally share a common URI infix.

Attribute clustering blocking~\cite{DBLP:journals/tkde/PapadakisIPNN13} exploits schematic information of the descriptions to minimize the number of unnecessary comparisons. To achieve this, prior to token blocking, it clusters attributes based on the similarities of their values over the entire entity collection. 
Each attribute from one collection is connected to its most similar attribute in the other collection and connected attributes, taken by transitive closure, form non-overlapping clusters. 
Then, 
each token $t$ in the values of an attribute, belonging to a cluster $c$, defines a block $b_{c.t}$. Hence, comparisons between descriptions without a common token in a similar attribute, are discarded. 
Like token blocking, attribute clustering generates overlapping blocks. Compared to the blocks of token blocking, it produces a larger number of smaller blocks. 

\begin{figure}[b]
\begin{center}
\includegraphics[scale=0.42]{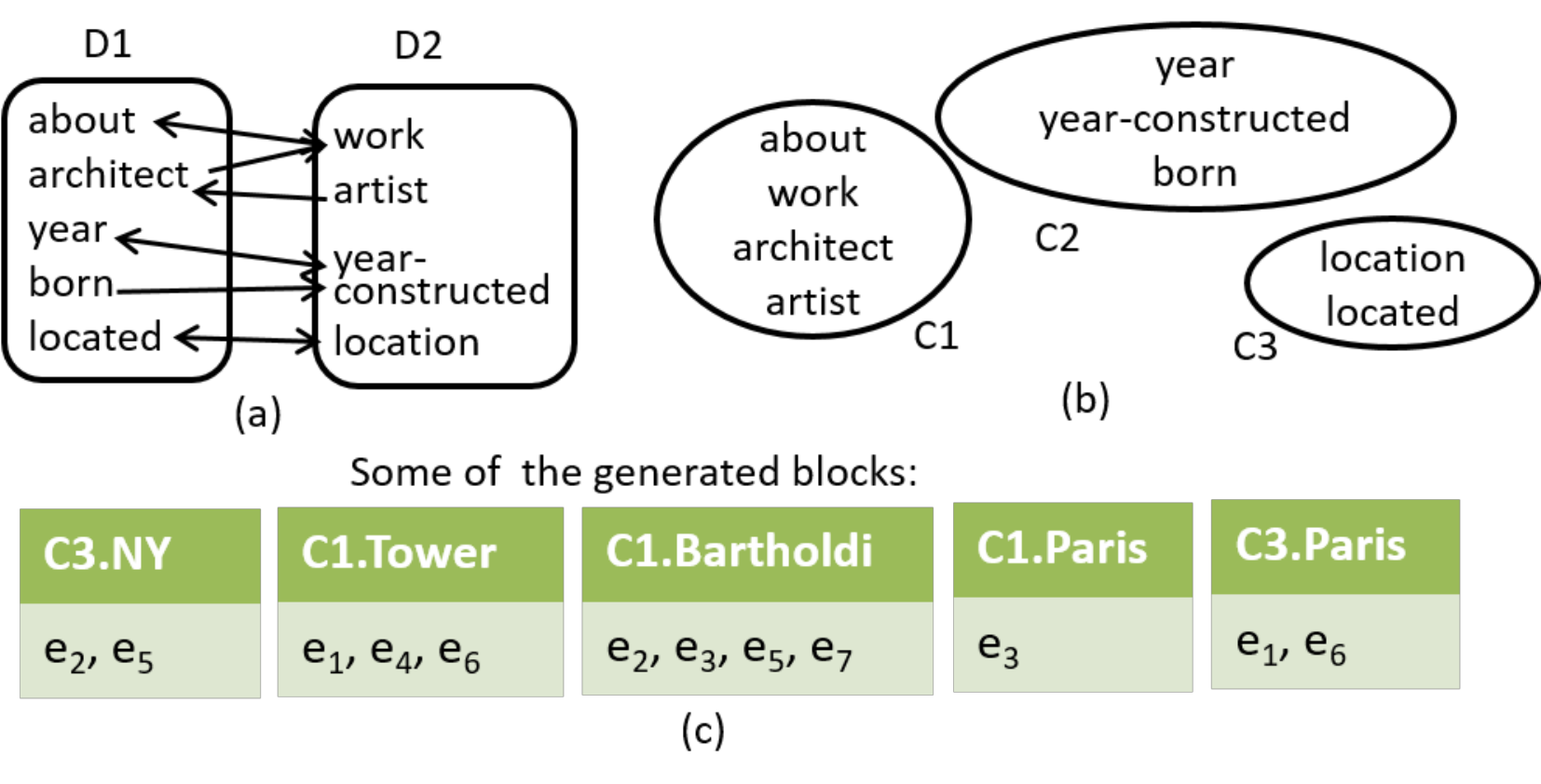} 
\vspace{-11pt}
\caption{Attribute clustering blocking example. Pairs of most similar attributes are linked (a). Connected attributes form clusters (b). Descriptions with a common token in the values of attributes of the same cluster, are placed in a common block (c).}
\label{fig:attribute_ex}
\vspace{-15pt}
\end{center}
\end{figure}

As an example, consider that the descriptions of Figure \ref{fig:descriptions} consist of two clean collections, $D_1 = \{e_1, e_2, e_3, e_4\}$ and $D_2 = \{e_5, e_6, e_7\}$. 
Using Jaccard similarity, the attribute $work$ (with values: $\{$Lady, Liberty, Eiffel, Tower, Bartholdi, Fountain$\}$) of $D_2$ is the most similar attribute to $about$ of $D_1$. 
Similarly, the transitive closure of the pairs of most similar attributes between $D_1$ and $D_2$ (Figure~\ref{fig:attribute_ex}(a) depicts such pairs), produce the clusters of attribute names (Figure~\ref{fig:attribute_ex}(b)). 
A subset of the blocks constructed for each cluster is shown in Figure~\ref{fig:attribute_ex}(c). This way, the comparisons ($e_1, e_3$) and ($e_3, e_6$) that were suggested by token blocking, due to the common token \emph{Paris}, are now discarded, since the token \emph{Paris} appears in different attribute clusters for $e_3$ than for $e_1$ and $e_6$, as shown in the bottom blocks of Figure~\ref{fig:attribute_ex}(c). 
Again, both unnecessary (e.g., $e_4$ and $e_6$ are both placed in block $C1.Tower$ (Figure~\ref{fig:attribute_ex}(c))), and redundant (e.g., ($e_1, e_3$) is still contained in 4 different blocks) comparisons are generated. 

Unlike previous methods analyzing the content of descriptions, prefix-infix(-suffix) blocking \cite{DBLP:conf/wsdm/PapadakisINPN12} exploits the naming pattern in the descriptions' URIs.  
The prefix describes the domain of the URI, the infix is a local identifier, and the optional suffix contains details about the format, or a named anchor. For example, the prefix of ``http://liris.cnrs.fr/olivier.aubert/foaf.rdf\#me'' is ``http://liris.cnrs.fr'', the infix is ``/olivier.aubert'' and the suffix is ``/foaf.rdf\#me''. 
Given a set of descriptions, 
this method creates one block for each token in the descriptions literal values and one block for each URI infix. 
It is constrained by the extent to which common naming policies are followed by the KBs. In a favourable scenario, it creates additional blocks than token blocking for the names of the descriptions, which enables to consider matching descriptions, even with no common tokens in their literal values.  

\begin{figure}
\begin{center}
\includegraphics[scale=0.42]{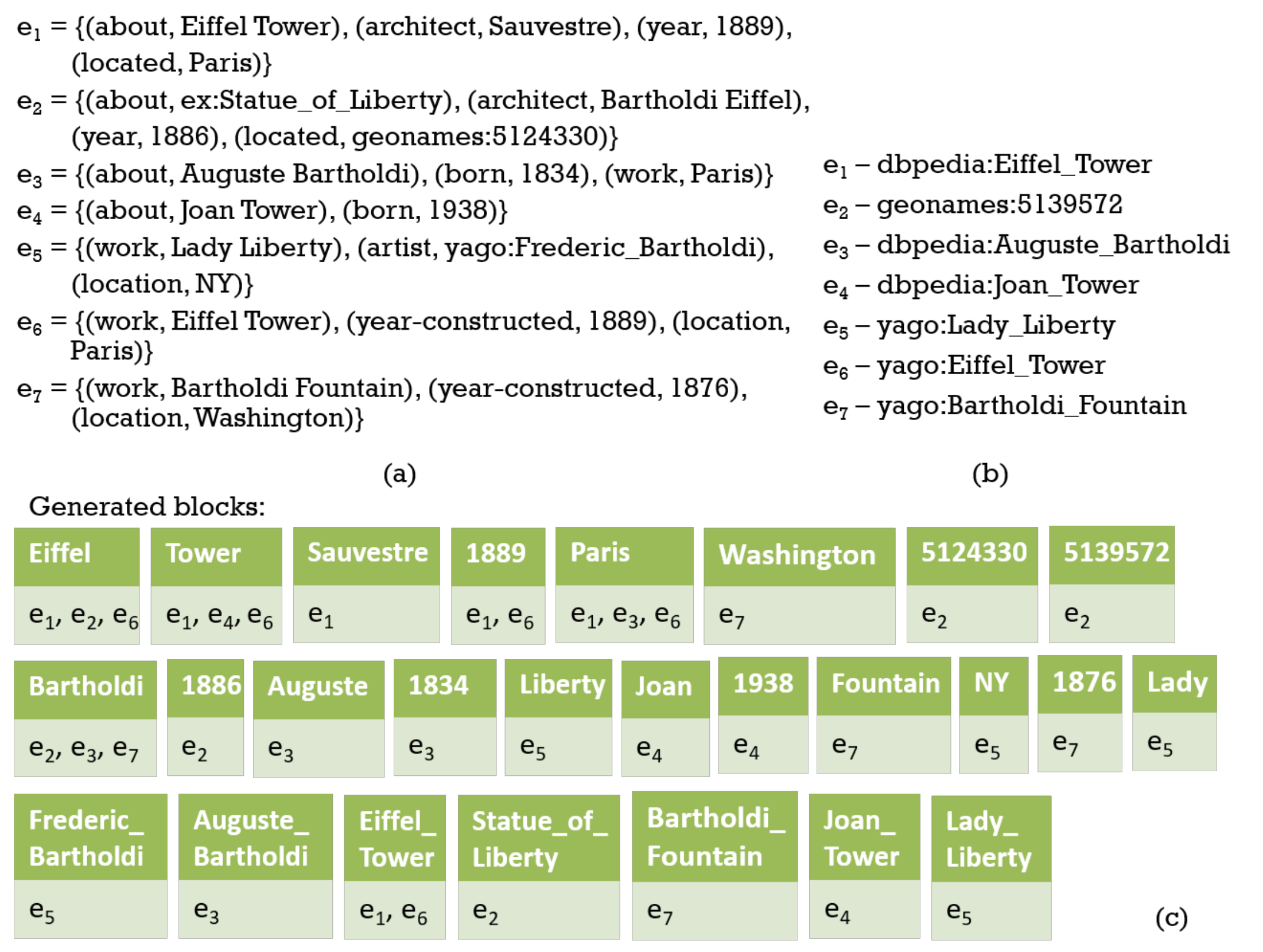} 
\vspace{-11pt}
\caption{Prefix-infix(-suffix) blocking example. A set of descriptions (a), their subject URIs (b), and the blocks from their tokens and infixes (c).}
\label{fig:prefix_ex}
\vspace{-15pt}
\end{center}
\end{figure}

Figure~\ref{fig:prefix_ex}(c) shows the blocks produced after applying prefix-infix(-suffix) blocking to the descriptions of Figure~\ref{fig:prefix_ex}(a) (the descriptions of Figure~\ref{fig:descriptions}, slightly modified to illustrate the characteristics of the method), while Figure~\ref{fig:prefix_ex}(b) presents the URI identifiers of the descriptions.

\begin{figure}[b]
\begin{center}
\includegraphics[scale=0.42]{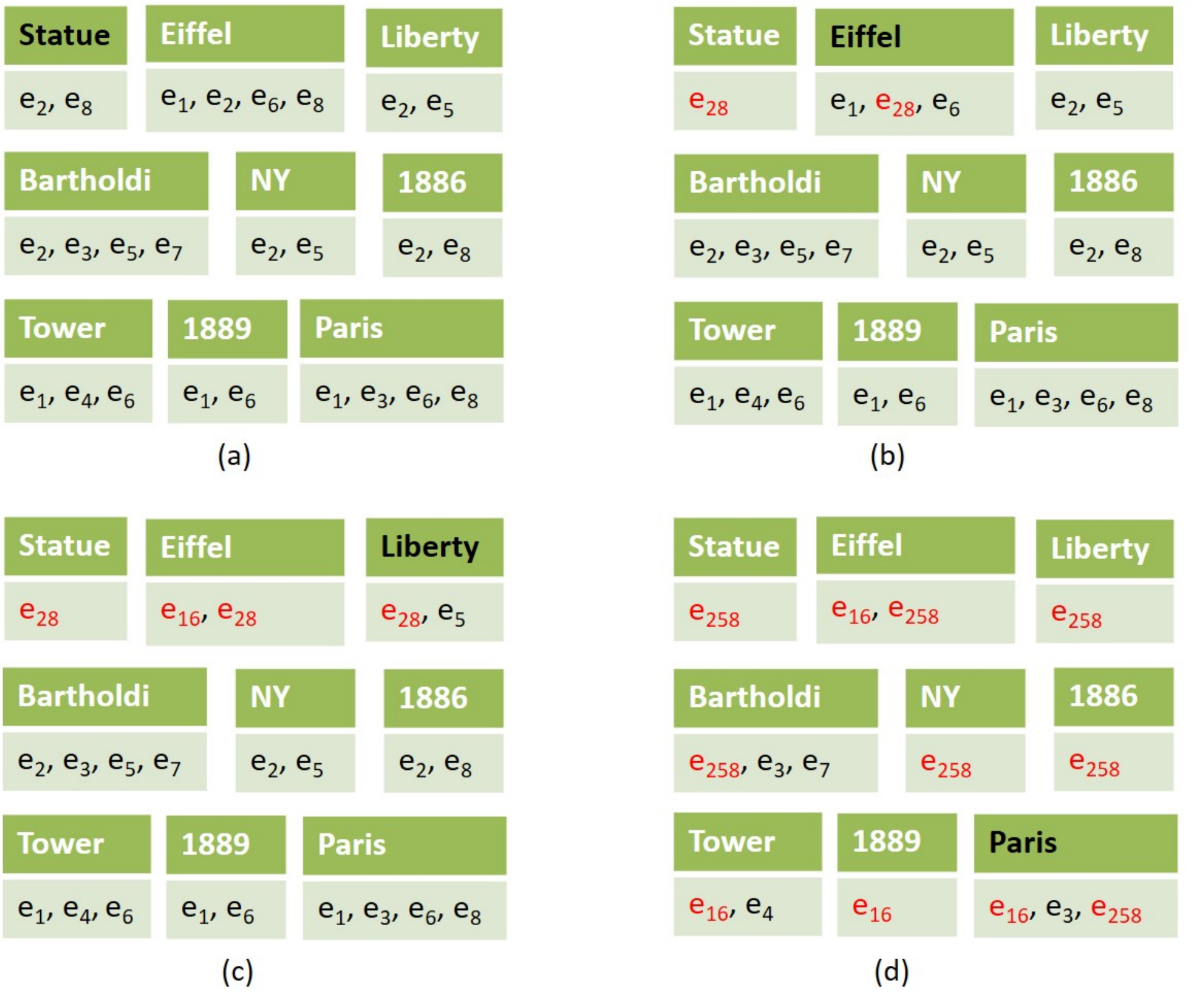} 
\vspace{-11pt}
\caption{Iterative blocking example. Given a set of blocks (a), the matches of each block are merged, propagating the results to the subsequent blocks (b),(c), until no more merges are possible (d).}
\label{fig:iterative_blocking_ex.jpg}
\vspace{-15pt}
\end{center}
\end{figure}

Recent works have proposed using an iterative ER process, interleaved with blocking. The intuition  is that the ER results of a processed block, may help identifying more matches in another block. Specifically, iterative blocking~\cite{DBLP:conf/sigmod/WhangMKTG09}, applied on the results of blocking, examines one block at a time looking for matches. When a match is found in a block, the resulting merging of the descriptions that match is propagated to all other blocks, replacing the initial matching descriptions. This way, redundant comparisons between the same pair of descriptions in different blocks are saved and, in addition, more matches can be identified efficiently. The same block may be processed multiple times, until no new matches are found. Inherently, iterative blocking follows a sequential execution model, since the results of ER in one block directly affect other blocks.

Given the results of token blocking for the dirty collection of Figure~\ref{fig:descriptions} and an additional description $e_8$ = \{(work, Statue of Lib.), (architect, Eiffel), (year-constructed, 1886)\}, the process of iterative blocking is presented in Figure~\ref{fig:iterative_blocking_ex.jpg}. 
The results of token blocking, excluding blocks with a single description, are shown in Figure~\ref{fig:iterative_blocking_ex.jpg}~(a). 
Starting from the block $Statue$, ($e_2, e_8$) is found to be matching. The matching descriptions are merged into a new description $e_{28}$,  representing both. Hence, in the next step (Figure~\ref{fig:iterative_blocking_ex.jpg}~(b)),
$e_2$ and $e_8$ are replaced by $e_{28}$ in block $Eiffel$, so they are not compared again. Also, another match ($e_1, e_6$) is found in this block and the result of the merging $e_{16}$ is replacing the descriptions $e_1$ and $e_6$ in the following step ((Figure~\ref{fig:iterative_blocking_ex.jpg}~(c))). Also in this step, $e_{28}$ is compared to $e_5$ and they are found to match, creating the new description $e_{258}$. Note that this match could not be identified by the other blocking methods, since $e_5$ and $e_8$ did not share any common blocks. Finally, when no new merges occur in an iteration, after processing all the blocks, the algorithm terminates, yielding the results $P = \{\{e_{16}\}, \{e_{258}\}, \{e_3\}, \{e_4\},$ $\{e_7\}\}$, as shown in Figure~\ref{fig:iterative_blocking_ex.jpg}~(d). Those are the results of ER and not blocking.

Overall, Table~\ref{tab:criteria} summarizes, simplified, the criteria employed by the aforementioned blocking methods to place
two descriptions into the same block. In this context, the following questions naturally arise:

\begin{compactitem}
\item Which blocking method performs best and for which dataset characteristics? 
\item How could we evaluate a blocking method, and its ability to reduce the number of suggested comparisons versus its ability to correctly place matching descriptions in common blocks? 
\item Does blocking place all matches in common blocks and what are the characteristics of the missed matches?
\item Could blocking be used for identifying other types of relations, e.g., geographical, between descriptions? 
\end{compactitem}

\begin{table}
\scriptsize
\centering
\caption{Criteria for placing descriptions in the same block.}
\vspace{-11pt}
\begin{tabular}{|p{0.72in}|p{2.45in}|}
\hline
{\bf Method} & {\bf Criterion}  \\ \hline
\hline
{\em Token Blocking} & The descriptions have a common token in their values.\\ \hline
{\em Attribute Clustering Blocking} & The descriptions have a common token in the values of attributes that have similar values in overall.\\ \hline
{\em Prefix-Infix(-Suffix) Blocking} & The descriptions have a common token in their literal values, or a common URI infix.  \\ \hline
\multirow{3}{*}{{\em Iterative Blocking}} & The descriptions are placed in a block by a blocking method, or after replacing descriptions that were merged in a previous iteration.\\ \hline
\end{tabular}
\label{tab:criteria}
\vspace{-11pt}
\end{table}

%% file: implementation.tex
\section{MapReduce Implementation}
\label{sec:implementation}
Next, we present the MapReduce version of the evaluated methods, designed to cope with Web data\footnote{Source code and datasets available at \url{csd.uoc.gr/~vefthym/minoanER/}.}. 
Note that iterative blocking~\cite{DBLP:conf/sigmod/WhangMKTG09} is an inherently sequential process, hence, we do not provide an implementation for it in MapReduce.

\subsection{Token Blocking} \label{ssec:TokenBlockingMR}
\begin{figure}[t]
\begin{center}
\includegraphics[scale=0.35]{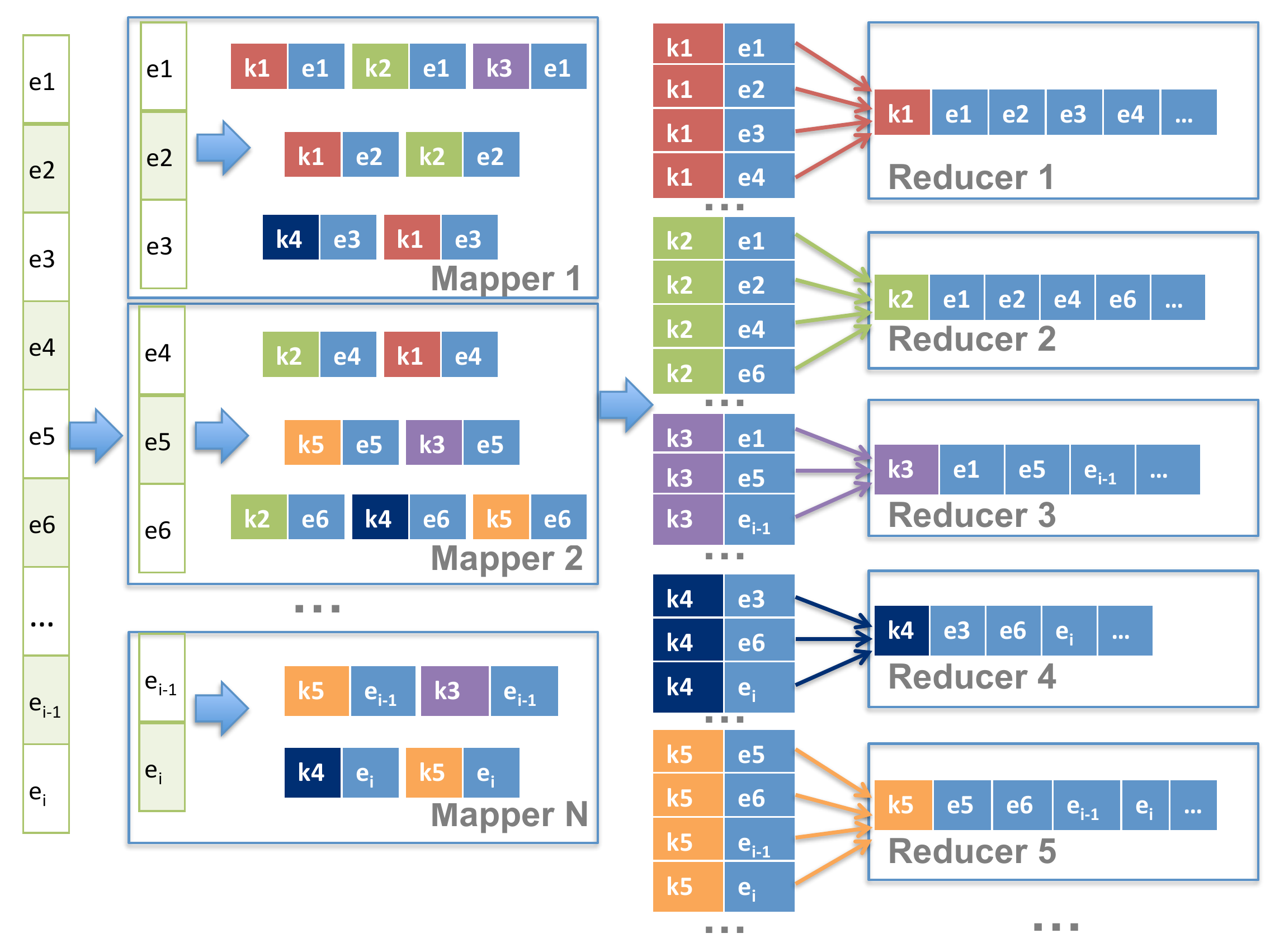} 
\vspace{-11pt}
\caption{Token blocking in MapReduce.}
\label{fig:token}
\vspace{-15pt}
\end{center}
\end{figure}

Token blocking is essentially an inverted index of descriptions. Each token is a key in this index, associated with a list of all the descriptions containing it. 
Our implementation of token blocking in MapReduce is based on the procedure illustrated in Figure~\ref{fig:token}.
In the map phase, one entity description of the local input split is processed at a time. For each token $t$ in the values of a description $e_i$, a ($t$, $e_i$) pair is emitted by the mapper. 
In the reduce phase, all descriptions having a common token will be processed by the same reduce function, i.e., placed in the same block.

\subsection{Attribute Clustering Blocking} \label{ssec:AttributeClusteringMR}

\begin{figure*}[t]
\begin{center}
\includegraphics[scale=0.35]{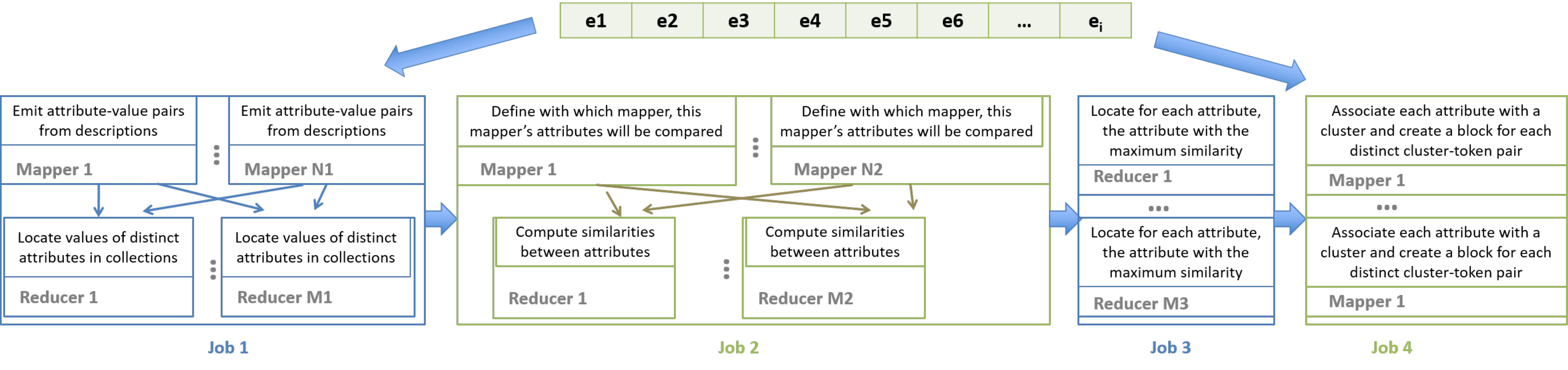} 
\vspace{-15pt}
\caption{Attribute clustering blocking in MapReduce.}
\label{fig:attribute}
\vspace{-15pt}
\end{center}
\end{figure*}

Given two clean entity collections, our implementation of attribute clustering blocking can be briefly sketched by the following steps, each representing a MapReduce job. 
Figure~\ref{fig:attribute} illustrates a high-level flow of the process.

\textbf{Attribute Creation.} First, we gather the values of each attribute. In the map phase, we emit an (\textit{attribute}, \textit{value}) pair for each attribute-value pair in a description. We also keep the collection of this attribute in the \textit{key}. In the reduce phase, all the values of an attribute are grouped together and their concatenation is emitted as the value of this attribute.

\textbf{Attribute Similarities.} In the second job, we compute the pairwise Jaccard similarities between the trigram sets of all attributes. 
A mapper outputs each input attribute, as many times, as the number of total mappers. Each time, a composite \textit{key}, consisting of the current mapper id and another mapper id, will determine in which reducer the attribute will be placed, and to which other attributes it will be compared. 
For example, assuming 3 mappers in total, the mapper with id 2, emits for each input attribute, 3 different \textit{keys}: $1\_2$, $2\_2$, and $2\_3$. 
The keys $1\_2$ and $2\_3$ will result in comparing the contents of mapper 2 to the contents of mappers 1 and 3, while $2\_2$ will result in comparing the contents of mapper 2 to each other. 
The \textit{value} of each emitted pair is the input attribute with its values and the current mapper id. 
In the reduce phase, we compute similarities of attributes, ensuring that each comparison is performed once. 
For each pair of attributes, we emit a (\textit{key}, \textit{value}) pair, with one attribute being the \textit{key} and the second attribute along with their similarity score being the \textit{value}.

\textbf{Best Match.} In the third job, we use an identity mapper, which just forwards its input. A combiner keeps for each attribute of each collection, only the attribute of the other collection with the local highest similarity score. 
In the reduce phase, we pick for each attribute of each collection, the attribute with the maximum similarity score, in overall, from the other collection. 
Before this job ends, we start the first step of clustering the most similar attributes together. 
To accomplish that, we emit for each best-matching attribute pair, two (attribute, clusterId) pairs, one for each attribute, with the same clusterId. Ids of clusters with common attributes are marked, in order to be merged at the next step.

\textbf{Final Clustering and Blocking.} In the final job, we associate each attribute with a final cluster id, according to the marks of the previous step. 
Then, we perform token blocking (Section \ref{ssec:TokenBlockingMR}), with only difference that in each \textit{key} emitted from a mapper, there is also a cluster prefix, enabling distinctions between blocks for the same token. For example, if the same token $t$ appears in a description $e_i$ for attributes in clusters $c_j$ and $c_k$, then the mapper will emit the pairs $(c_j.t, e_i)$ and $(c_k.t, e_i)$, instead of a single $(t, e_i)$.

\subsection{Prefix-Infix(-Suffix)} \label{PrefixInfixMR}
\begin{figure*}[t]
\begin{center}
\includegraphics[scale=0.35]{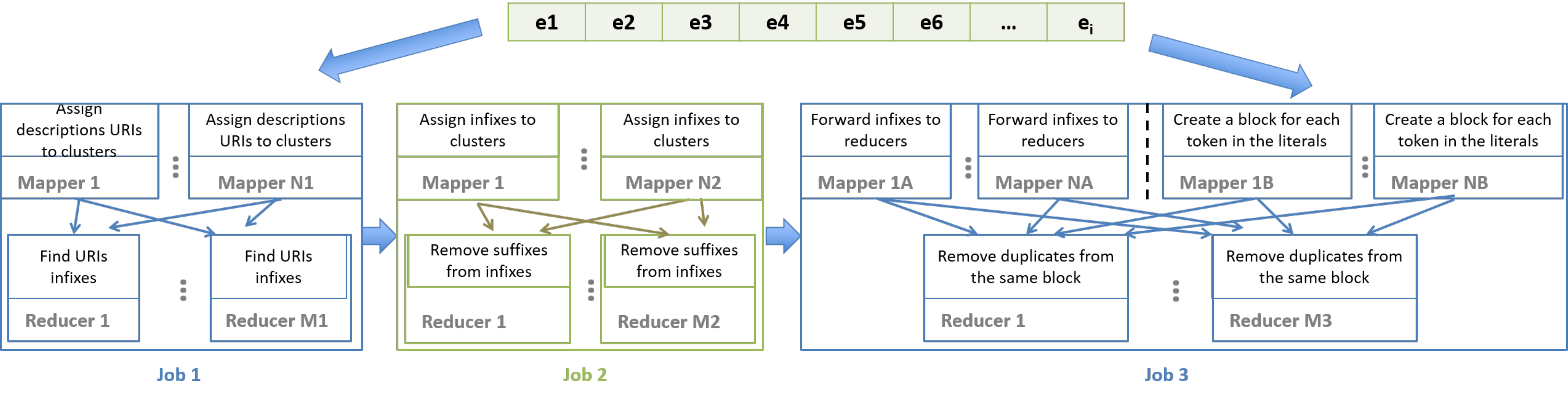} 
\vspace{-15pt}
\caption{Prefix-infix(-suffix) blocking in MapReduce.}
\label{fig:prefix}
\vspace{-15pt}
\end{center}
\end{figure*}

Our MapReduce implementation of this method consists of three jobs.
The first two are the MapReduce adaptation of the infix extraction algorithm   \cite{DBLP:conf/wsdm/PapadakisINPN12}. The third job reads the descriptions, as well as the infixes produced by the second job and creates the blocks. 
A high-level representation of the process is depicted in Figure~\ref{fig:prefix}.

\textbf{Prefix Removal.} 
In the map phase, we output a (\textit{key}, \textit{value}) pair for each URI in a description. 
The \textit{key} is the second token of the URI (after ``http'') and the \textit{value} consists of the whole URI and the identifier of the entity description having this URI. This clusters the URIs according to their second token, which usually represents the domain (e.g., ``dbpedia''), in the reduce phase. 
For each URI in a cluster, we find, among all its possible prefixes, the one with the largest set of distinct (immediately) next tokens. The part of the URI following the prefix is the \textit{key} of each output pair, with \textit{value} consisting of the input \textit{key}, i.e., the second token of the URI, and the entity identifier having this URI.

\textbf{Suffix Removal.} We apply Prefix Removal, on each reverse URI (without prefix), to remove the suffix. 

\textbf{Infix\&Token Blocking.} We create the final blocks, based on the output of Suffix Removal and the initial entity collection. We use two different mappers, operating in parallel; an identity mapper, forwarding the output of Suffix Removal and the mapper of token blocking, operating on the tokens of literal values only of the input descriptions. 
In the reduce phase, all the descriptions having a common token or infix in their literals or URIs will be placed in the same block.

%% file: experiments_setup.tex
\section{Experimental Setup}
\label{sec:setup}

In this section, we present the experimental framework we have designed for evaluating existing blocking algorithms. We describe the datasets and the measures we employed to study the behavior of the blocking algorithms under different characteristics of entity descriptions in the LOD cloud. 
We have used a cluster of 15 Ubuntu 12.04.3 LTS servers (1 master, 14 slaves), each with 8 CPUs, 8GB RAM and 60GB of disk, provided by {\raise.17ex\hbox{$\scriptstyle\sim$}}okeanos~\cite{DBLP:journals/internet/KoukisVK13}. Each node could run simultaneously 4 map or reduce tasks, each with a heap size of 1250MB, leaving resources required for I/O and communication with the master.
We used Apache Hadoop 1.2.0 and Java version 1.7.0\_25 from OpenJDK. 

\subsection{Datasets}\label{ssec:Datasets}
Our study relies on real data from the Billion Triples Challenge 2012 dataset\footnote{\url{km.aifb.kit.edu/projects/btc-2012/}} (BTC12), DBpedia, Kasabi\footnote{\url{archive.org/details/kasabi}}, the Linked Archives Hub project\footnote{\url{data.archiveshub.ac.uk/}}, and OAEI benchmarks\footnote{\url{oaei.ontologymatching.org/}}. 
To capture the differences in the heterogeneity and semantic relationships of descriptions, we distinguish between data originating from sources in the {\em center} and the {\em periphery} of the LOD cloud. In general, central sources, such as DBpedia and Freebase, are derived from a common source, Wikipedia, from which they extract information regarding an entity. Such descriptions often refer to the original wiki page and feature synonym attributes whose values share a significant number of common tokens. Since they have been exhaustively studied in the literature, descriptions across central LOD sources are heavily interlinked using in their majority {\em owl:sameAs} links \cite{DBLP:conf/semweb/SchmachtenbergBP14}, expressing equivalence relations. 
In our experiments, we used the DBpedia (\textit{BTC12DBpedia}) and Freebase (\textit{BTC12Freebase}) datasets from BTC12, and the raw infoboxes from DBpedia 3.5 (\textit{Infoboxes}), i.e., two different versions of DBpedia. 
From the OAEI benchmarks, we used the one including the \textit{DBLP} and \textit{Rexa} (OAEI 2009) - describing authors and publications - 
dataset, that has been widely used in the literature (e.g., in~\cite{DBLP:conf/kdd/Lacoste-JulienPDKGG13}).
We also included a movies dataset
, used in \cite{DBLP:journals/tkde/PapadakisIPNN13}, extracted from DBpedia movies and \textit{IMDB}, to validate the correctness of our algorithms.

On the other hand, data sources in the periphery of the LOD cloud are far more diverse to each other and sparsely interlinked. 
In our experiments, we considered the \textit{BTC12Rest}, the \textit{BBCmusic} and the \textit{LOCAH} datasets. \textit{BTC12Rest} originates from the BTC12 dataset, which consists of multiple data sources, like DBLP, geonames and drugbank. \textit{BBCmusic} originates from Kasabi and contains descriptions regarding music bands and artists, extracted from MusicBrainz and Wikipedia. For \textit{LOCAH}, we used the latest published version at Archives hub (March 2014). This, rather small dataset links descriptions of people, from UK archival institutions, with their descriptions in DBpedia. 

\input{datasetsTable}

Table~\ref{tab:datasets2} provides statistics about these datasets, for the number of contained triples, descriptions, attributes, and the average number of attribute-value pairs per description. We have also included the number of entity types, taken as the distinct values of the property \textit{rdf:type}, when provided. 
Observe that \textit{BTC12DBpedia} contains more types than attributes. This is due to the fact that DBpedia entities may have multiple types from taxonomic ontologies like Yago. \textit{IMDB} is the dataset with the highest number of attribute-value pairs per description. Finally, we have included in each dataset the number of duplicate descriptions based on our ground truth, i.e., descriptions that have been reported to be equivalent (via {\em owl:sameAs} links) across all datasets of our testbed. Taking into account the transitivity of equality, those descriptions should be regarded as matches, too.

To investigate the ability of blocking algorithms in recognizing relatedness links beyond the \textit{owl:sameAs} among descriptions, we considered the Kasabi \textit{airports} and \textit{airlines} datasets, containing data linked to DBpedia, the dataset with the highest number of references, with the \textit{umbel:isLike} property. This property is used to associate entities that may or may not be equivalent, but are believed to be so. 
The \textit{twitter} dataset contains data for the presentations of an ESWC conference. It is linked to DBpedia with the \textit{dct:subject} property, which captures relatedness of entities to topics and it is also used in the \textit{books} and \textit{iati} datasets. \textit{Books} describes books listed in the English language section of Dutch printed book auction catalogues of collections of scholars and religious ministers from the $17^{th}$ century. \textit{Iati} contains data from the International Aid Transparency Initiative. \textit{Iati} is also connected to DBpedia with the \textit{dct:coverage} property, which associates an entity to its spatial or temporal topic, its spatial applicability, or the jurisdiction under which it is relevant. Finally, the \textit{www2012} dataset contains data from the WWW2012 conference, linked to DBpedia with the \textit{foaf:based\_near} property, which associates an entity to an abstract notion of location. 
Table~\ref{tab:otherDatasets} details the type and the number of links of these datasets to DBpedia.

\input{otherDatasetsTable}

In this setting, we combine \textit{BTC12DBpedia} with each of the datasets of Table~\ref{tab:datasets2} to produce the entity collections presented in Table~\ref{tab:collections}, on which we finally ran our experiments. 
To combine two datasets, for the dirty ER setting, we simply concatenate them into a singe file, while for clean-clean ER, we seek candidate matches between those datasets.

\noindent - $\boldsymbol{D1}$ combines \textit{BTC12DBpedia} with \textit{Infoboxes}. Since it contains two versions of the same dataset, it is considered as a homogeneous collection. This is the biggest collection in terms of triples, as well as attributes.

\noindent - $\boldsymbol{D2}$ combines \textit{BTC12DBpedia} with \textit{BTC12Rest}. Since it is constructed by many different datasets, it is the most heterogeneous collection. Note that \textit{BTC12Rest} has the highest number of attributes per entity type.

\noindent - $\boldsymbol{D3}$ combines \textit{BTC12DBpedia} with \textit{BTC12Freebase}. It is the biggest collection in terms of entity descriptions, matches, entity types and comparisons.

\noindent - $\boldsymbol{D4}$ combines \textit{BTC12DBpedia} with \textit{BBCmusic}. Note that \textit{BBCmusic} extracts some of its data from MusicBrainz, which, in turn, extracts data from Wikipedia. Also, \textit{BBCmusic} is edited and maintained by users and BBC staff.

\noindent - $\boldsymbol{D5}$ combines \textit{BTC12DBpedia} with \textit{LOCAH}, the smallest dataset, both in terms of triples and entity descriptions.

\noindent - $\boldsymbol{D6}$ combines DBpedia movies and \textit{IMDB}, as originally used in \cite{DBLP:journals/tkde/PapadakisIPNN13}. It is the most homogeneous collection, it only contains descriptions of movies (i.e., a single entity type) using the smallest number of attributes among all collections. However, the significantly greater (even by six orders of magnitude, compared to the other collections) ratio of matches to non-matches  is not typical of the collections we can find in the Web of data. 

\noindent - $\boldsymbol{D7}$ combines \textit{DBLP} and \textit{Rexa}. Both datasets use the same ontology; \textit{Rexa}'s attributes are a subset of those used by \textit{DBLP}. Also, it is the collection with the lowest number of attribute-value pairs per description. Note that this collection is a typical benchmark used to evaluate instance matching algorithms.

\input{collectionsTable}

Following the distinction of our datasets between central and peripheral, we also distinguish our collections between central ($D1$, $D3$, $D6$, and $D7$), composed of central datasets, and peripheral ($D2$, $D4$, and $D5$), part of which are peripheral datasets.
For all the collections, we consider both their \textit{clean-clean} and \textit{dirty} versions. In practice, for our datasets, the \textit{clean-clean} and \textit{dirty} versions of a collection are the same; their distinction serves only as means for measuring how well a blocking method can identify links across different datasets and within the same dataset. We finally combine \textit{BTC12DBpedia} with each peripheral dataset of Table~\ref{tab:otherDatasets} to produce entity collections for studying the ability of blocking algorithms to discover different relatedness attributes.

{\bf GroundTruth.}
Our ground truths were built using a methodology met in the literature (e.g.,~\cite{DBLP:conf/wsdm/PapadakisINPN12,DBLP:journals/tkde/PapadakisIPNN13}). For $D2$-$D5$, we consider the \textit{owl:sameAs} links to/from DBpedia 3.7 (the version used in BTC12). For $D1$, we consider the subject URIs of \textit{Infoboxes} that also appear as subjects in \textit{BTC12DBpedia}. The ground truth of $D6$, provided in~\cite{DBLP:journals/tkde/PapadakisIPNN13}, is made of DBpedia movies connected with IMDB movies through the \textit{imdbId} property. The ground truth of $D7$ is provided by OAEI, since it is a benchmark collection, containing equivalence links between authors, as well as publications.
Based on the ground truth and the generated blocks, we say that a known matching pair of descriptions is correctly resolved, i.e., a true positive (TP), if there is at least a block, to which both these descriptions belong. Pairs belonging to the same block are candidate matches. A false positive (FP) is a distinct candidate match not contained in the ground truth. In the opposite, if a known pair of matching descriptions is not a candidate match, this pair is considered a false negative (FN). All remaining pairs of descriptions are considered to be true negatives (TN). 

Similarly to $D2$-$D5$, we used the available types of links of the datasets of Table~\ref{tab:otherDatasets} to \textit{BTC12DBpedia}, instead of \textit{owl:sameAs}, to produce the ground truth of the corresponding entity collections. From all datasets, except $D6$ and $D7$, we removed the triples present in the ground truth, since identifying those links is the goal of our tasks.

Our pre-processing, implemented in MapReduce, parses RDF triples in order to transform them into entity descriptions, which are the input of the methods used in our study. It simply groups the triples by subject, and outputs each group as an entity description, using the subject as the entity identifier, removing triples containing a blank node. Moreover, we kept only the entity descriptions for which we know their linked description in \textit{BTC12DBpedia} and removed the rest. This way, we know that any suggested comparison between a pair of descriptions outside the ground-truth is false. 

\begin{table}[t]
\vspace{-5pt}
\scriptsize
\centering
\caption{Quality Measures.}
\vspace{-11pt}
\begin{tabular}{|p{1.15cm}|p{3.05cm}|p{3.45cm}|}
\hline
{\bf Name} & {\bf Formula} & {\bf Description} \\ \hline
\hline
\multirow{3}{*}{Recall} & \multirow{3}{*}{$\frac{TP}{TP + FN}$} & Measure what fraction of the known matches are candidate matches. \\ \hline
\multirow{3}{*}{Precision} & \multirow{3}{*}{$\frac{TP}{TP + FP}$} & Measure what fraction of the candidate matches are known matches. \\ \hline
\multirow{2}{*}{F-measure} & \multirow{2}{*}{$2\frac{Precision \cdot Recall}{Precision + Recall}$}  & The harmonic mean of precision and recall. \\ \hline
\multirow{3}{*}{$RR$} &\multirow{3}{*}{$1-\frac{\text{comparisons with blocking}}{\text{comparisons without blocking}}$} & Returns the ratio of reduced comparisons when blocking is applied. \\ \hline 
\multirow{2}{*}{$H3R$} & \multirow{2}{*}{$2\frac{RR \cdot Recall}{RR + Recall}$} & The harmonic mean of recall and reduction ratio. \\ \hline
\end{tabular}
\label{tab:measures}
\vspace{-11pt}
\end{table}

\subsection{Measures}\label{ssec:Measures}
The employed quality measures along with a short description are summarized in Table~\ref{tab:measures}. The range of all measures is $[0,1]$, with 1 being the ideal value. 
The recall of a blocking method is the upper recall threshold of a non-iterative ER algorithm
, which takes its generated blocks as input. 
Therefore, (1-recall) represents the \textit{cost of blocking}. 
$RR$ is the percentage of comparisons that we save if we apply the given blocking method. Consequently, it reflects the \textit{benefit of blocking}, since the reason for using blocking in the first place, is the reduction in the required comparisons. 

In general, a good blocking method should have a low impact on recall, i.e., a low cost, and a great impact on the number of required comparisons, i.e., a high benefit.  
Typically, this trade-off is captured by the F-measure, the harmonic mean of recall and precision. However, as we will see in the next section, the values of F-measure are dominated by the values of precision, which are many orders of magnitude lower than those of recall, so F-measure cannot be easily used to express this trade-off. Moreover, precision is not as important as recall is for blocking, since precision can  only be improved by a non-iterative ER method that follows blocking, whereas the recall of blocking is the upper threshold of such ER methods. 
Thus, we define $H3R$ as the harmonic mean of recall and reduction ratio, a measure which has also been used in~\cite{DBLP:conf/icdm/KejriwalM13}. Similar to the F-measure, $H3R$ gives high values only when both recall and reduction ratio have high values. Unlike F-measure, $H3R$ manages to capture the trade-off between effectiveness and efficiency in a more balanced way.  
Note that $H3R$ does not estimate the performance of a blocking approach (as, for example, \cite{DBLP:conf/wsdm/PapadakisINPN12} does), but evaluates it  based on the actual results. 

%% file: datasetsTable.tex
\begin{table*}[ht!]
\scriptsize
\centering
\caption{Datasets characteristics.}
\vspace{-11pt}
\begin{tabular}{|p{2.0cm}|c|x{1.6cm}|x{2cm}|c|c|x{1.8cm}|c|}
\hline
 &\multirow{3}{*}{{\footnotesize {\bf RDF triples}}} & {\footnotesize {\bf entity descriptions}} & {\footnotesize {\bf avg. attribute-\newline value pairs per description}} &  \multirow{3}{*}{{\footnotesize {\bf attributes}}} & \multirow{3}{*}{{\footnotesize {\bf entity types}}} &{\footnotesize  {\bf attributes/ entity types}} & \multirow{3}{*}{{\footnotesize {\bf duplicates}}}
\\ \hline 

BTC12DBpedia & 102,306,242 & 8,945,920 & 11.44 & 36,354 & 258,202 & 0.14 & 0\\ \hline
Infoboxes & 27,011,880 & 1,638,149 & 16.49 & 31,857 & 5,535 & 5.76 & 0\\ \hline
BTC12Rest & 849,656 & 31,668 & 26.83 & 518 & 33 & 15.7 & 863\\ \hline
BTC12Freebase & 25,050,970 & 1,849,180 & 13.55 & 8,323 & 8,232 & 1.01 & 12,058\\ \hline
BBCmusic & 268,759 & 25,359 & 10.60 & 29 & 4 & 7.25 & 372\\ \hline
LOCAH & 12,932 & 1,233 & 10.49 & 14 & 4 & 3.5 & 250\\ \hline
DBpedia$_{mov}$ & 180,680 & 27,615 & 6.54 & 5 & 1 & 5 & 0\\ \hline
IMDB & 816,012 & 23,182 & 35.20 & 7 & 1 & 7 & 0\\ \hline
DBLP & 12,074,269 & 1,642,945 & 7.35 & 30 & 10 & 3 & 0 \\ \hline
Rexa & 64,787 & 14,771 & 4.39 & 12 & 3 & 4 & 0 \\ \hline

\end{tabular}
\label{tab:datasets2}
\end{table*}

%% file: otherDatasetsTable.tex
\begin{table}
\vspace{-11pt}
\scriptsize
\centering
\caption{Characteristics of datasets with different types of links to \textit{BTC12DBpedia}.}
\vspace{-11pt}
\begin{tabular}{|p{0.88cm}|x{0.8cm}|x{1.1cm}|x{1.5cm}|x{1.32cm}|x{0.6cm}|}
\hline
& \multirow{4}{0.8cm}{{\footnotesize {\bf RDF triples}}} & {\footnotesize {\bf entity descriptions}}  & {\footnotesize {\bf avg. att.-value pairs per description}} & \multirow{4}{*}{{\footnotesize {\bf link}}} & \multirow{4}{*}{{\footnotesize {\bf links}}} \\ \hline
airports &238,973&12,294&19.44&\textit{umbel:isLike}&12,269 \\ \hline
airlines &15,465&1,141&13.55&\textit{umbel:isLike}&1,217 \\ \hline
twitter &6,743&2,932&2.30&\textit{dct:subject}&20,671 \\ \hline
books &2,993&748&4.00&\textit{dct:subject}&1,605 \\ \hline
\multirow{2}{*}{iati} & \multirow{2}{*}{378,130}&\multirow{2}{*}{31,868}&\multirow{2}{*}{11.87}&\textit{dct:subject}&23,763 \\ \cline{5-6}
&&&&\textit{dct:coverage}&7,833 \\ \hline
www2012 &11,772&1,547&7.61&\textit{foaf:based\_near}&1,562 \\ \hline
\end{tabular}
\label{tab:otherDatasets}
\vspace{-11pt}
\end{table}


%% file: collectionsTable.tex
\begin{table*}[h!t]
\scriptsize
\centering
\caption{Entity collections characteristics.}
\vspace{-11pt}
\begin{tabular}{|p{3.8cm}|c|c|c|c|c|c|c|}
\hline
 & {\footnotesize {\bf D1}} & {\footnotesize {\bf D2}} & {\footnotesize {\bf D3}} & {\footnotesize {\bf D4}} & {\footnotesize {\bf D5}} & {\footnotesize {\bf D6}} & {\footnotesize {\bf D7}} \\ \hline
\hline
RDF triples & 129,318,122 & 103,155,898 & 127,357,212 & 102,575,001 & 102,319,174 & 996,692 & 12,139,056 \\ \hline
entity descriptions & 10,584,069 & 8,977,588 & 10,795,100 & 8,971,279 & 8,947,153 & 50,797 & 1,657,716\\ \hline
avg. attribute-value pairs &\multirow{2}{*}{12.22} & \multirow{2}{*}{11.49} & \multirow{2}{*}{11.80} & \multirow{2}{*}{11.43} & \multirow{2}{*}{11.44} & \multirow{2}{*}{19.62} & \multirow{2}{*}{7.32} \\ 
per description  & & & & & & & \\ \hline
attributes & 68,211 & 36,872 & 44,677 & 36,383 & 36,368 & 12 & 42 \\ \hline
entity types & 263,737 & 258,232 & 266,434 & 258,206  & 258,205 & 1 & 10 \\ \hline
\hline
matches &1,564,311&30,864&1,688,606&23,572&1,087&22,405&1,532 \\ \hline
matches (incl. duplicates) &1,564,311&31,727&1,700,664&23,944&1,337&22,405&1,532 \\ \hline
matches/non-matches &$1.07 \cdot 10^{-7}$&$1.09 \cdot 10^{-7}$&$1.02 \cdot 10^{-7}$&$1.04 \cdot 10^{-7}$&$9.85 \cdot 10^{-8}$&$3.5 \cdot 10^{-5}$& $6.3 \cdot 10^{-8}$ \\ \hline
matches/non-matches (dirty) & $2.79 \cdot 10^{-8}$ & $7.87 \cdot 10^{-10}$ & $2.92 \cdot 10^{-8}$ & $5.95 \cdot 10^{-10}$ &$3.34 \cdot 10^{-11}$ & $1.74 \cdot 10^{-5}$&$1.1 \cdot 10^{-9}$ \\ \hline
comparisons {\tiny (w/o blocking)} & \multicolumn{7}{l|}{} \\ \cline{2-8}
~ ~ ~ {\em clean-clean} & $1.47 \cdot 10^{13}$&$2.83 \cdot 10^{11}$&$1.65 \cdot 10^{13}$&$2.27 \cdot 10^{11}$ & $1.1 \cdot 10^{10}$ & $6.4 \cdot 10^8$ & $2.4 \cdot 10^{10}$ \\  \cline{2-8}
~ ~ ~ {\em dirty} & $5.6 \cdot 10^{13}$&$4.03 \cdot 10^{13}$&$5.83 \cdot 10^{13}$&$4.02 \cdot 10^{13}$ & $4 \cdot 10^{13}$ & $1.29 \cdot 10^9$ & $1.37 \cdot 10^{12}$ \\ \hline
\end{tabular}
\label{tab:collections}
\vspace{-11pt}
\end{table*}

%% file: experiments_evaluation.tex
\section{Experimental Evaluation}
\label{sec:experiments}
In this section, we analyze the quality and performance of the evaluated blocking methods, taking into consideration the specific features of each dataset. Then, we present the results of blocking, when different kinds of links, other than \textit{owl:sameAs}, are used as ground truth and conclude with a discussion of the lessons learned from this analysis. 
In our evaluation, we use the adaptation of  token blocking $ToB$, attribute clustering $AtC$ and prefix-infix(-suffix) blocking $PIS$ in MapReduce. Both $ToB$ and $PIS$ can be used with either \textit{clean-clean} $Cl$ or \textit{dirty} $Di$ entity collections, while $AtC$ is suitable for $Cl$ collections.
Moreover, the process of $AtC$ requires a similarity function; we use Jaccard similarity over the set of trigrams from the values (similar to \cite{DBLP:journals/tkde/PapadakisIPNN13}).

\subsection{Quality Results}\label{ssec:qualityResults}

\input{qualityTable}

\subsubsection{Identified Matches (TPs)}\label{sssec:recall}
\begin{figure*}[h!t]
\centering
\includegraphics[height=34mm]{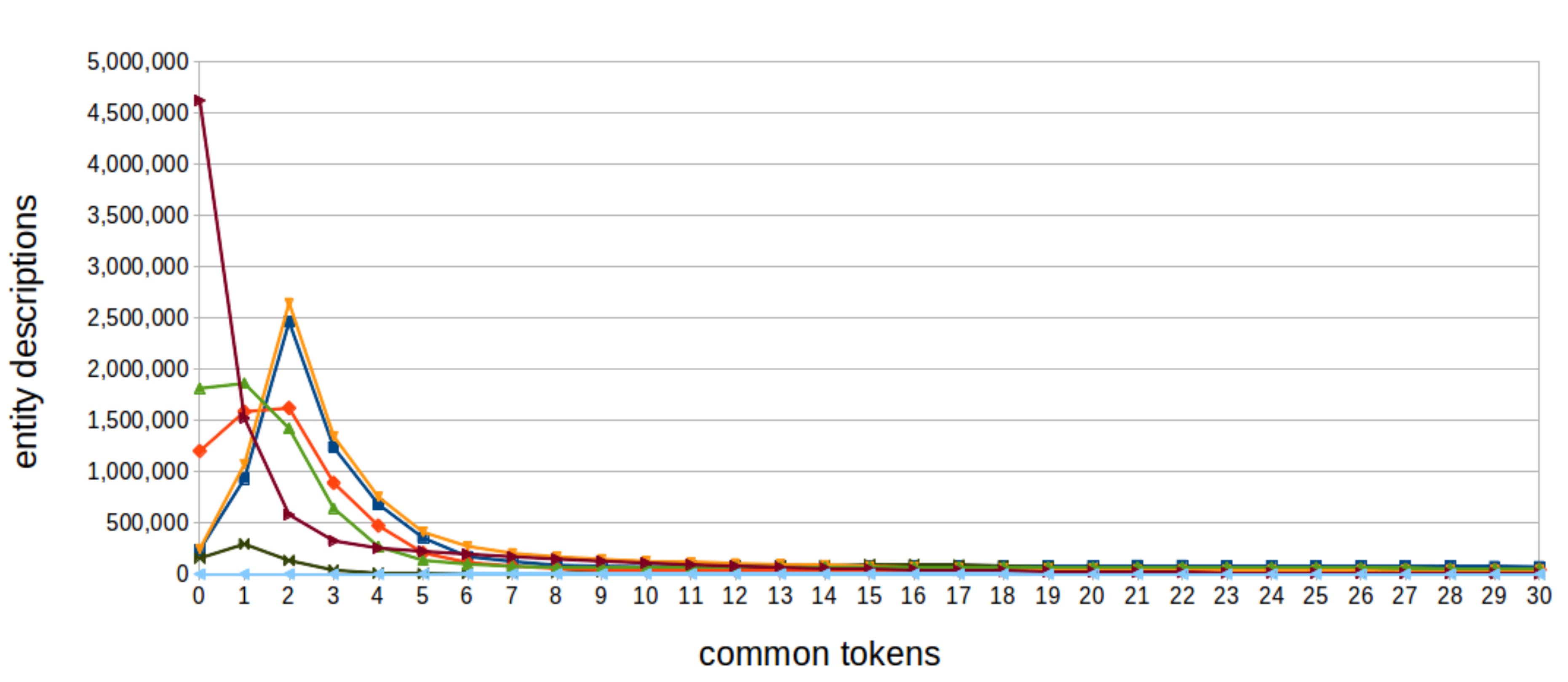}
\includegraphics[height=35mm]{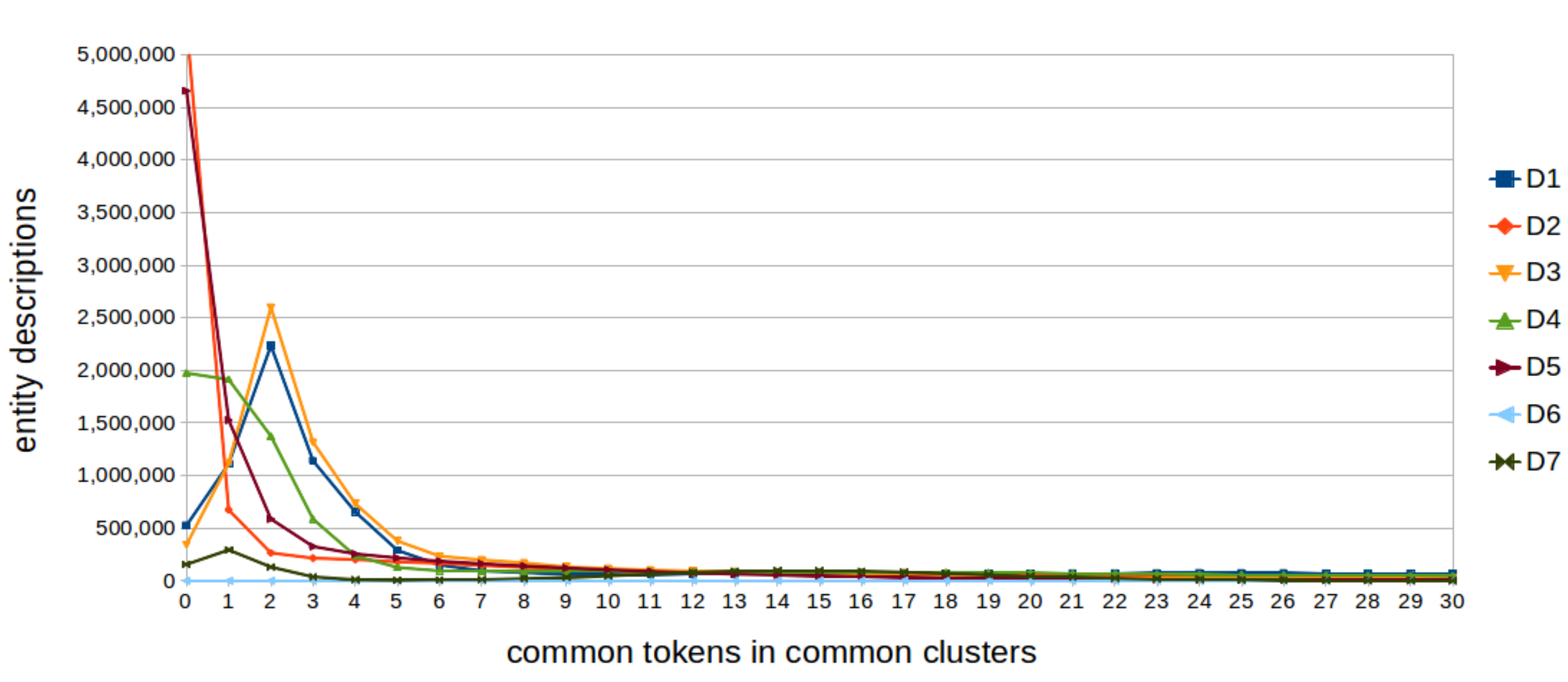}
\vspace{-11pt}
\caption{Common tokens (left) and common tokens in common clusters (right) per entity description distributions for $D1$-$D7$.}
\label{fig:commonTokensDistributions}
\vspace{-15pt}
\end{figure*}

\noindent{\bf Token blocking:}
The premise of this algorithm is that matching descriptions should at least share a common token, disregarding the comparisons between descriptions that do not share common tokens. Therefore, the higher the number of common tokens, i.e., tokens shared by the datasets composing an entity collection, a description has, the higher the chances it will be placed in a block with a matching description, increasing recall.
Figure~\ref{fig:commonTokensDistributions} (left) presents the distributions of common tokens per description, showing that descriptions in central collections feature many more common tokens than those in peripheral ones\footnote{We take the median values and not the averages, as the latter are highly influenced by extreme values and our distributions are skewed.}. For example, $41.43\%$ and $44\%$ of descriptions in $D1$ and $D3$, respectively, have 2-4 common tokens, while for $D2$, $D4$ and $D5$ the corresponding values are $33.26\%$, $26.03\%$ and $12.97\%$. We observe a big difference in the distributions of $D6$ and $D7$, which contain many more common tokens per description, to those of the other collections. 
Only $23.75\%$ of the descriptions in $D6$ and $44\%$ of the descriptions in $D7$ have 0 - 10 common tokens. Figure~\ref{fig:commonTokensDistributions} (left) also shows that a big number of descriptions in peripheral collections, do not share any common tokens. Those are hints that the recall of token blocking in central collections is higher than in peripheral collections. 

Indeed, $D6$ is the dataset with the highest recall ($99.92\%$) and the highest number of common tokens per entity (19), while $D5$ is the dataset with the lowest recall ($72.13\%$) and number of common tokens per entity (0). 
There is a big difference in the number of common tokens in $D6$, compared to $D1$ and $D3$, which is not reflected by their small difference in recall. 
Due to the high ratio of matches to non-matches in $D6$ (Table~\ref{tab:collections}), descriptions in this collection have many common tokens and this leads to high recall.

\noindent{\bf Attribute clustering blocking:}
The goal of attribute clustering is to improve the precision of token blocking, while retaining its recall as much as possible (it cannot have higher recall).
To do this, it restricts the number of attributes on which descriptions, featuring a common token, should be compared. 
Comparisons between descriptions that do not share a common token in a common attribute cluster, are discarded.
Hence, descriptions with many common tokens in common clusters are more likely to be matched.
Figure~\ref{fig:commonTokensDistributions}~(right) presents the distributions of the number of common tokens in common attribute clusters per entity. It shows a clearer distinction between central and peripheral collections than Figure~\ref{fig:commonTokensDistributions}~(left); the descriptions in central collections have many more common tokens in common clusters, while many descriptions in peripheral collections do not have any common token in a common cluster. 
This occurs, because values in the descriptions of peripheral collections are much less similar than those of central collections, leading to a bad clustering of the attributes and, thus, to lower recall.
In fact, $D6$ is the dataset with the highest recall ($99.55\%$) and the highest number of common tokens in common attribute clusters per entity ($19$). On the other hand, $D2$ and $D5$, which have the lowest recall values ($68.42\%$ and $71.11\%$) also have the lowest number of common token in common attribute clusters per entity ($0$).

In central collections ($D1$, $D3$, $D6$, $D7$), many, small clusters of similar attributes are formed, as the values of the descriptions are similar. This leads to a minor (or zero, in $D7$) decrease in recall, compared to token blocking, while it significantly improves its precision (even by an order of magnitude in $D3$).
$D1$ forms many (16,886), small attribute clusters (of 2 attributes in the median case), since in most cases there is a 1-1 mapping between the attributes of the datasets that compose it. These clusters contain the same attribute used by the two versions of DBpedia.

However, this approach has a substantial impact on recall in peripheral collections ($D2$, $D4$, $D5$), even if it still improves precision in all collections (even by an order of magnitude for $D4$). 
The descriptions in those collections have few common tokens, in the first place, which leads to a bad clustering of attributes; few clusters of many attributes, not similar to each other, are formed.
Hence, if we make the blocking criterion of token blocking stricter, by also considering attributes, then the more distinct attributes used per entity type, the more difficult it is for an entity description, to be placed in a common block with a matching description. 
For \textit{BTC12Rest} (part of $D2$), the ratio between attributes and entity types (last row of Table~\ref{tab:datasets2}) is the highest (15.7), leading to a great impact on recall (-$24.04\%$). 
This dataset has the biggest number of data sources that compose it and many different attribute names can be used for the same purpose; hence, big attribute clusters are formed.
\textit{LOCAH} (part of $D5$) only has 3.5 attributes per entity type. Thus, the recall of attribute clustering blocking is insignificantly reduced (-$1.02\%$), compared to that of token blocking.

\noindent{\bf Prefix-Infix(-Suffix) blocking:}
Prefix-Infix(-Suffix) blocking is built on the premise that many URIs contain useful information. Its goal is to extend token blocking and improve both its recall, by also considering the subject URIs of the descriptions, and its precision, by disregarding some unneeded tokens in the URI values (either in the prefix or suffix). It achieves good recall values in datasets with similar naming policies in the URIs, as in $D4$, part of which is \textit{BBCmusic}, which also has Wikipedia as a source. However, it misses many matching pairs of descriptions, when the names of the URIs do not contain useful information, as in $D3$ that uses random strings as ids, or have different policies, as in $D5$, which uses concatenations of tokens, without delimiters, as URIs. The recall of $D1$ is $100\%$, because the collection is constructed this way; it consists of two versions of the same dataset, DBpedia, and the URIs appearing as subjects in \textit{Infoboxes} are only those URIs that also appear as subjects in \textit{BTC12DBpedia}.
$PIS$ is not applicable (marked N/A) to $D6$, since URIs have been replaced with numerical ids in the provided datasets. 
In $D7$, recall is the same as in the other blocking methods, since the matches can be found by tokens in the literal values of the descriptions.



\subsubsection{Missed Matches (FNs)}\label{sssec:fns}
A non-negligible number of matching pairs of descriptions do not share any common tokens at all. 
Such descriptions, constituting the false negatives of token blocking, should not be assumed faulty, or noisy. 
We distinguish two different sources of information that can be exploited for successfully placing descriptions of missed matches in common blocks: 
\begin{compactenum}[(i)]
\item The matches of their neighbors: Given that a description can have, as one of its values, another description, neighborhoods of related descriptions are formed, spinning the Web of data. 
The knowledge of matches in the neighbors of a description is valuable for correctly matching this description. For example, if a description $e_{10}$ is related to $e_1$, $e_{20}$ is related to $e_2$, and we know that $e_{10}$ and $e_{20}$ match, then we can use this knowledge as a hint that $e_1$ and $e_2$ could possibly match, too. 
\item A third, matching description: In dirty collections (typically peripheral), which are composed of datasets that  potentially contain duplicate descriptions, a description $e_1$ could have more than one matching description, e.g., both $e_2$ and $e_3$. Identifying one of these matches, e.g., ($e_1$, $e_3$), knowing that ($e_2$, $e_3$) is a match, leads to also identify the missing match ($e_1$, $e_2$). 
\end{compactenum}
Table~\ref{tab:FNs} provides details about the number and the characteristics of false negative pairs of descriptions, and the set of individual descriptions that constitute these pairs\footnote{$D6$ is excluded, as it does not contain any descriptions with neighbors and $D7$ is excluded, as it only yields 7 missed matches.}.

We focus first on the neighbors of these descriptions, namely descriptions that appear in their values. 
We found that almost all the descriptions in the false negatives have at least one neighbor (second row of Table~\ref{tab:FNs}). 
Looking more thoroughly, we counted the percentage of descriptions in false negatives that have at least one neighbor belonging to the ground truth (third row of Table~\ref{tab:FNs}). In all cases, this percentage is more than $10\%$ and goes up to $58\%$ for $D4$. 
This means that, not only do these descriptions have neighbors, but many of these neighbors can be matched to other descriptions in the same collection as well. 
Then, we counted the percentage of descriptions in false negatives that have neighbors, which have already been matched to another description (fourth row of Table~\ref{tab:FNs}). This percentage is over $20\%$ in most collections, while it reaches up to $51.84\%$ for $D4$. 
Finally, we counted the percentage of false negative pairs, whose descriptions have neighbors, which match to each other (fifth row of Table~\ref{tab:FNs}). 
This percentage is 0 for $D1$, as matches in this collection are defined as descriptions that have the same subject URI. 
However, in some peripheral collections ($D2$, $D4$), examining the matches of the neighbors of the descriptions is meaningful.

Another useful piece of information for the missed matches of dirty collections is whether their descriptions have been correctly matched to a third description.
The last row of Table~\ref{tab:FNs} quantifies this statistic, showing that there are collections, both peripheral ($D2$, $D5$) and central ($D3$), for which this kind of information could, indeed, be useful.

The information of Table~\ref{tab:FNs} is lost when we only consider the tokens in the values of the descriptions to create the blocks in a single round, but it could be useful to an iterative method. Iterative blocking~\cite{DBLP:conf/sigmod/WhangMKTG09}, based on some initial blocks, aims to identify matches of type (ii), as well as eliminate redundant comparisons. 
In our experiments, the recall of iterative blocking, given the blocks of token blocking from the dirty collection with the smallest number of comparisons ($D6$), was the same as that of token blocking ($99.92\%$), since both of its datasets contain no duplicates (Tables~\ref{tab:datasets2},~\ref{tab:collections}), but the number of comparisons performed was almost half of those suggested by token blocking. 
We also applied iterative blocking to the dirty collection with the lowest recall ($D5$), giving the blocks generated by token blocking as input. The process did not terminate within a reasonable amount of time, 
even so, the recall of iterative blocking was $78.09\%$ after a first pass, whereas the recall of token blocking was $77.34\%$.  
\begin{table}
\scriptsize
\centering
\caption{Characteristics of the missed matches of token blocking.}
\vspace{-11pt}
\begin{tabular}{|p{1.1in}|x{0.27in}|x{0.27in}|x{0.27in}|x{0.27in}|x{0.27in}|x{0.27in}|}
\hline
 & \textbf{D1} & \textbf{D2} & \textbf{D3} & \textbf{D4} & \textbf{D5} \\ \hline \hline
FNs &25,419&	3,176&	87,672&	2,886&	303 \\ \hline
descriptions in FNs, with  neighbor(s)&\multirow{2}{*}{$99.64\%$}&\multirow{2}{*}{$100\%$}&\multirow{2}{*}{$99.99\%$}&\multirow{2}{*}{$100\%$}&\multirow{2}{*}{$100\%$} \\ \hline
descriptions in FNs, with neighbor(s) in ground truth &\multirow{3}{*}{$22.60\%$}&\multirow{3}{*}{$53.94\%$}&\multirow{3}{*}{$36.43\%$}&\multirow{3}{*}{$58.36\%$}&\multirow{3}{*}{$11.57\%$} \\ \hline
descriptions in FNs, with neighbor(s) with an identified match &\multirow{3}{*}{$20.94\%$}&\multirow{3}{*}{$48.54\%$}&\multirow{3}{*}{$34.05\%$}&\multirow{3}{*}{$51.84\%$}&\multirow{3}{*}{$7.59\%$} \\ \hline
FNs with matching neighbors &\multirow{2}{*}{$0\%$}&\multirow{2}{*}{$24.81\%$}&\multirow{2}{*}{$0.38\%$}&\multirow{2}{*}{$37.63\%$}&\multirow{2}{*}{$0\%$}\\ \hline
FNs with common, identified matches&\multirow{2}{*}{$0\%$}&\multirow{2}{*}{$25.35\%$}&\multirow{2}{*}{$10.54\%$}&\multirow{2}{*}{$0.14\%$}&\multirow{2}{*}{$8.58\%$}\\ \hline
\end{tabular}
\label{tab:FNs}
\vspace{-11pt}
\end{table}

Regarding attribute clustering blocking, it misses the matches that are also missed by token blocking, plus matches that, even if they share common tokens, those tokens appear in the values of attributes in different clusters. 
The matches missed by prefix-infix(-suffix) blocking are those with no common tokens in their literal values and no common infixes in their URIs.

\subsubsection{Non-matches (FPs and TNs)}\label{sssec:negative}
Next, we examine the ability of blocking methods to identify non-matches, namely their ability to avoid placing non-matching descriptions in the same block. A key statistic for this, regarding the datasets, is the ratio of matches to non-matches (Table~\ref{tab:collections}). The higher the ratio, the easier it is for a blocking method to have better precision, as it statistically has better chances of suggesting a correct comparison. $D6$ is the collection with the highest such ratio and precision, while $D5$ has the lowest ratio and, in most blocking methods, the lowest precision, too.
It is clear from Table~\ref{tab:recall} that attribute clustering is the most precise method, since, in almost every case, it results in the fewest wrong suggestions. On the contrary, the least precise method is token blocking, in all cases. The differences in precision, in some cases even by an order of magnitude, also determine F-measure, since the differences in recall are not that big. All the evaluated methods have very low precision, i.e., the vast majority of suggested comparisons correspond to non-matches.

\subsubsection{Structural Analysis of Matches and Non-matches}\label{sssec:analysis}
To better understand the characteristics of matches versus those of non-matches in the evaluated collections, we have analyzed sample pairs of matching and non-matching descriptions. In particular, we have taken 1,000 random pairs of matches and non-matches from each collection and we have focused on their neighbor pairs of descriptions. The results of this analysis are presented in Table~\ref{tab:analysis}. 

First, we counted the number of pairs of descriptions that both have neighbors. We found that those numbers, presented in the first two rows of Table~\ref{tab:analysis} for matches and non-matches, respectively, are almost the same. Practically, almost all the pairs of descriptions are linked to other pairs of description, in all collections. Then, we measured the median number of neighbors (pairs of descriptions) that a match has (Table~\ref{tab:analysis}, row 3) and the same median number for non-matches (Table~\ref{tab:analysis}, row 4). Again, there are no significant differences between those two lines. Those numbers vary greatly from collection to collection, ranging from 1 (for $D7$) to 171 (for $D4$). Finally, we counted the number of pairs, whose neighbor pairs match. For matches (Table~\ref{tab:analysis}, row 5), this number is always higher than the corresponding number for non-matches (Table~\ref{tab:analysis}, row 6). Intuitively, this means that when a match is found, the chances that there is another match in its neighbor pairs are increased. 

\begin{table}
\scriptsize
\centering
\caption{Analysis of 1K sampled matches and 1K sampled non-matches.}
\vspace{-11pt}
\begin{tabular}{|p{0.8in}|x{0.25in}|x{0.25in}|x{0.25in}|x{0.25in}|x{0.25in}|x{0.25in}|x{0.25in}|}
\hline
 & \textbf{D1} & \textbf{D2} & \textbf{D3} & \textbf{D4} & \textbf{D5} & \textbf{D7} \\ \hline \hline
matches with neighbors & \multirow{2}{*}{967} & \multirow{2}{*}{956} & \multirow{2}{*}{913} & \multirow{2}{*}{918} & \multirow{2}{*}{859} & \multirow{2}{*}{973} \\ \hline
non-matches with neighbors & \multirow{2}{*}{966} & \multirow{2}{*}{955} & \multirow{2}{*}{912} & \multirow{2}{*}{917} & \multirow{2}{*}{854} & \multirow{2}{*}{973} \\ \hline
neighbors of \newline matches (median) &  \multirow{3}{*}{17} &  \multirow{3}{*}{80} & \multirow{3}{*}{100} & \multirow{3}{*}{138} & \multirow{3}{*}{121} & \multirow{3}{*}{1}  \\ \hline
neighbors of non-matches (median) &  \multirow{3}{*}{72} &  \multirow{3}{*}{80} & \multirow{3}{*}{105} & \multirow{3}{*}{171} & \multirow{3}{*}{121} & \multirow{3}{*}{1}   \\ \hline
matches with matching neighbors	& \multirow{3}{*}{862} & \multirow{3}{*}{254} &  \multirow{3}{*}{7} & \multirow{3}{*}{766} & \multirow{3}{*}{570} & \multirow{3}{*}{966} \\ \hline
non-matches with matching neighbors	&  \multirow{3}{*}{32} &  \multirow{3}{*}{22} & \multirow{3}{*}{0} &   \multirow{3}{*}{0} & \multirow{3}{*}{542} & \multirow{3}{*}{590} \\ \hline

\end{tabular}
\label{tab:analysis}
\vspace{-11pt}
\end{table}

\subsection{Performance Results}\label{ssec:performance}

Table~\ref{tab:recall} shows that all the evaluated methods manage to greatly reduce the number of comparisons that would be required if blocking was not employed, e.g., by one ($D1$-$D4$, $D7$) or two ($D5$) orders of magnitude for token blocking. This is reflected by high $RR$ in all cases. An exception is $D6$, which is much smaller in terms of descriptions and, consequently, comparisons without blocking. Moreover, its descriptions contain many more common tokens than the other collections, leading to more comparisons per entity. Therefore, token blocking does not save many of the comparisons that would be required without blocking and in $D6$ dirty, it even produces twice as many comparisons.

With respect to $H3R$\hide{and $H3R_2$}, we notice that, in general, central collections have higher scores, i.e., they present a better balance between recall and reduction ratio. This means that in these collections, comparisons that are discarded by blocking mostly correspond to non-matches, while many of the comparisons discarded by blocking in peripheral collections correspond to matches. 
Again, $D6$ has a different behaviour, since it initially contains a much smaller number of comparisons and a high ratio of matches to non-matches, so the reduction ratio for this collection is limited. These measures are not applicable to token blocking, when applied to $D6$ dirty, since in that case the reduction ratio is negative.

\begin{table*}
\scriptsize
\centering
\caption{Recall of the collections composed of datasets of Table~\ref{tab:otherDatasets} and
 \textit{BTC12DBpedia}.}
\vspace{-11pt}
\begin{tabular}{|l|c|c|c|c|c|c|c|}
\hline
 & \textbf{airports} & \textbf{airlines}  & \textbf{twitter} & \textbf{books} & \multicolumn{2}{c|}{\textbf{iati}} & \textbf{www2012} \\ \hline
\hline
link & \textit{umbel:isLike} & \textit{umbel:isLike} & \textit{dct:subject} & \textit{dct:subject}  & \textit{dct:subject} & \textit{dct:coverage} & \textit{foaf:based\_near}\\ \hline
Recall of token blocking& $97.47\%$ & $99.75\%$ & $9.52\%$ & $63.55\%$ & $49.13\%$ & $39.46\%$ & $62.61\%$ \\ \hline
\end{tabular}
\label{tab:recallOfOtherDatasets}
\vspace{-11pt}
\end{table*}

\subsection{Different Types of Links}\label{ssec:otherLinksEval}
In order to evaluate the ability of blocking methods to identify more types of links, semantically close or even not that close to equivalence links, 
we have run a set of experiments with the peripheral collections consisting of each of the datasets of Table~\ref{tab:otherDatasets} and \textit{BTC12DBpedia}. 
We have chosen a wide range of link types, in order to show that the same blocking method can better identify some specific link types, and fail to identify other link types. 
Table~\ref{tab:recallOfOtherDatasets} provides the recall of token blocking, when applied to each of those collections.
Similarly to the \textit{owl:sameAs} links, token blocking performs well for links with the semantics of equivalence (i.e., \textit{umbel:isLike}, expressing a possible equivalence), as in the \textit{airports} and \textit{airlines} datasets with recall values close to $100\%$. It also manages to identify many subject associations (i.e., \textit{dct:subject}, expressing the topic of a description), as in the cases of \textit{books} and \textit{iati} datasets. It performs poorly in identifying this kind of association, however, in the \textit{twitter} dataset, where its recall values fall to below $10\%$. This could be justified by the nature of this dataset, which, in most cases, simply states who created a slideset. Regarding spatial associations (i.e., \textit{dct:coverage}, expressing the spatial topic of a description, and \textit{foaf:based\_near}, relating two spatial objects), token blocking manages to identify a mere $39\%$ of the \textit{coverage} associations of the \textit{iati} dataset, but it performs much better in identifying the \textit{based\_near} associations of \textit{www2012}, with a recall of $63\%$. The spatial relationships of \textit{coverage} are looser than those of \textit{based\_near}, hence the related entities are not so strongly related in the former type of links. For example, in \textit{iati}, the description of a project regarding the evaluation of cereal crop residues is linked to the DBpedia resource describing Latin America and the Caribbean, through the \textit{coverage} relation, while, in \textit{www2012}, a Greek professor is linked to the DBpedia resource describing Greece, through the \textit{based\_near} relation.

\subsection{Lessons Learned}\label{ssec:discussion}
We now present the key points of our evaluation.
\emph{Central} collections are mostly derived from Wikipedia, from which they extract information regarding an entity. 
This way, descriptions in such collections follow similar naming policies and feature many common tokens (Figure~\ref{fig:commonTokensDistributions}) in the values of semantically similar, or equivalent attributes (see the small size of clusters in Table~\ref{tab:recall}). Those are exactly the premises on which the evaluated blocking methods are built.

For these reasons, the recall achieved by token blocking in central entity collections is very high (ranges from $99.92\%$ to $94.85\%$). With the exception of $D6$ (featuring a higher ratio of matching to non-matching descriptions), the precision achieved by token blocking in these collections ranges from $2.49 \cdot 10^{-6}$ to $3.64 \cdot 10^{-7}$. The gains in precision brought by attribute clustering blocking in central entity collections are up to one order of magnitude (for $D3$), with a minor cost on recall (from $0\%$ to $3.42\%$). Prefix-infix(-suffix) blocking can improve both recall and precision of token blocking for central collections, as in $D1$, but, it can also deteriorate these values, as in the dirty case of $D3$, which uses random identifiers as URIs, in which recall drops by $7.79\%$ and precision by $26.72\%$.
In a nutshell, many redundant comparisons are suggested by blocking algorithms in all entity collections (see precision and F-measure in Table~\ref{tab:recall}), due to the small ratio of matches to non-matches in the collections (Table~\ref{tab:collections}). However, as $H3R$ reveals, the comparisons that are discarded by blocking in central collections mostly correspond to non-matches. 

On the contrary, descriptions in \emph{peripheral} KBs are more 
diverse, following different naming policies and sharing few common tokens (Figure~\ref{fig:commonTokensDistributions}), since they stem from various sources. The lack of similar values in those descriptions leads to a bad clustering of attributes; big clusters of attributes not similar to each other are formed (Table~\ref{tab:recall}).

For these reasons, the recall of token blocking for peripheral collections drops even to $72.13\%$, while precision ranges from $1.3 \cdot 10^{-6}$ to $1.29 \cdot 10^{-9}$. The gains in precision brought by attribute clustering blocking (up to one order of magnitude) in peripheral collections, come at the cost of a drop in recall up to $24.04\%$ (corresponding to 7,421 more missed matches). 
Prefix-infix(-suffix) blocking can improve the precision of token blocking in peripheral collections, even by an order of magnitude (for $D4$), or decrease it by an order of magnitude (for $D5$), while it decreases recall from $0.74\%$ to $3.96\%$, i.e., more matches are missed. In the case of $D4$, in which both datasets use Wikipedia as a source, recall is improved by up to $7.68\%$. In overall, however, $H3R$ reveals that many of the comparisons that are discarded by blocking in peripheral collections correspond to matches.

Nevertheless, information for the missed matches, e.g., from the neighborhoods of their descriptions (Table~\ref{tab:FNs}), sets the ground for a new generation of ER algorithms, which will exploit this information to identify more matches, in an iterative fashion. In Table~\ref{tab:analysis}, we have shown that even a single match in the neighborhood of a candidate pair is a good match-indication for that pair, too.

Finally, in peripheral collections, there are several types of relations, other than equivalence, between descriptions. Token blocking identifies some of them, depending on the dataset, the specific type of such links, and the immediacy of those relations (Table~\ref{tab:recallOfOtherDatasets}). It does not perform well when the data do not contain much information (e.g., see the characteristics of \textit{twitter} in Table~\ref{tab:otherDatasets}), or when the relationship of the entities is loose (e.g., see the recall of \textit{iati} for \textit{dct:coverage} in Table~\ref{tab:recallOfOtherDatasets}). 
Thus, for a quantitative evaluation of blocking methods ground truth should not be restricted only to {\em owl:sameAs} links.
We could potentially take other relations into account, to identify more such links, or more {\em owl:sameAs} links, using iterative algorithms.


%% file: qualityTable.tex
\begin{table*}[h!t]
\scriptsize
\centering
\caption{Statistics and evaluation of blocking methods.}
\vspace{-11pt}
\begin{tabular}{|p{4.2cm}|c|c|c|c|c|c|c|}
\cline{2-8}
\multicolumn{1}{l|}{} & \textbf{D1} & \textbf{D2}  & \textbf{D3} & \textbf{D4} & \textbf{D5} & \textbf{D6} & \textbf{D7} \\ 
\cline{2-8}

\multicolumn{8}{l}{Token blocking statistics:} \\ \hline
blocks & 1,639,962 & 122,340 & 1,019,501 & 57,085 & 2,109 & 40,304 & 18,553 \\ \hline
comparisons (clean-clean) & $1.68 \cdot 10^{12} $&$	3.74 \cdot 10^{10}	$&$6.56 \cdot 10^{11}$&$	2.39 \cdot 10^{10}$&$	8.72 \cdot 10^{8}$&$	2.91 \cdot 10^{8}$ & $1.45 \cdot 10^{9}$
 \\ \hline
RR (clean) & $88.51\%$&$86.81\%$&$96.03\%$&$89.48\%$&$92.09\%$&$54.50\%$ &$94.04\%$ \\ \hline
comparisons (dirty) & $5.56 \cdot 10^{12}$&$	3.68 \cdot 10^{12}$&$	4.27 \cdot 10^{12}$&$	4.02 \cdot 10^{12}$&$	1.01 \cdot 10^{12}$&$	2.05 \cdot 10^{9}$ & $2.35 \cdot 10^{11}$
 \\ \hline
RR (dirty) & $90.08\%$&$90.87\%$&$92.67\%$&$90.01\%$&$97.48\%$&$-58.85\%$ & $82.93\%$ \\ \hline \hline
common tokens per entity (median) & 4 & 3 & 4 & 2 & 0 & 19 & 12 \\ \hline

\multicolumn{8}{l}{Attribute clustering blocking statistics:} \\ \hline
blocks & 5,602,644 & 150,293 & 1,673,855 & 39,587 & 3,724 & 43,716 & 19,148  \\ \hline
comparisons & $3.22 \cdot 10^{11}$&$	4.20 \cdot 10^{9}$&$	1.84 \cdot 10^{11}$&$1.43 \cdot 10^{9}$&$	7.13 \cdot 10^{8}$&$	2.13 \cdot 10^{8}$ & $8.38 \cdot 10^{8}$ \\ \hline
RR & $97.80\%$&$98.52\%$&$98.89\%$&$99.37\%$&$93.54\%$&$66.80\%$ & $96.55\%$ \\ \hline \hline
common tokens in common att. clusters per entity (median) & \multirow{2}{*}{4} & \multirow{2}{*}{0} & \multirow{2}{*}{4} & \multirow{2}{*}{2} & \multirow{2}{*}{0} & \multirow{2}{*}{19} & \multirow{2}{*}{11} \\ \hline
attribute clusters & 16,886 & 124 & 2,106 & 6 & 8 & 4 & 8 \\ \hline
attributes per attribute cluster (median) & \multirow{2}{*}{2} & \multirow{2}{*}{142} & \multirow{2}{*}{9} & \multirow{2}{*}{4,261} & \multirow{2}{*}{3,946} & \multirow{2}{*}{3} & \multirow{2}{*}{3.5} \\ \hline

\multicolumn{8}{l}{Prefix-Infix(-Suffix) blocking statistics:} \\ \hline
blocks & 3,266,798 & 141,517 & 789,723 & 45,403 & 2,098 & N/A & 18,442 \\ \hline
comparisons (clean-clean) & $1.10 \cdot 10^{12}$&$	1.78 \cdot 10^{10}	$&$2.75 \cdot 10^{11}$&$	2.30 \cdot 10^{9}$&$	4.08 \cdot 10^{8}$ & N/A & $1.28 \cdot 10^{9}$ \\ \hline
RR (clean) & $92.48\%$&$93.72\%$&$98.34\%$&$98.99\%$&$96.30\%$ & N/A & $94.72\%$ \\ \hline
comparisons (dirty) & $4.39 \cdot 10^{12}$&$	3.45 \cdot 10^{12}$&$	5.34 \cdot 10^{12}	$&$3.32 \cdot 10^{12}$&$	1.76 \cdot 10^{12}$ & N/A & $2.23 \cdot 10^{11}$ \\ \hline
RR (dirty) & $92.16\%$&$91.44\%$&$90.84\%$&$91.76\%$&$95.59\%$ & N/A & $83.78\%$ \\ \hline

\multicolumn{8}{l}{Recall:} \\ \hline
Token blocking (clean-clean)&$98.38\%$&$92.46\%$&$95.52\%$&$87.76\%$&$72.13\%$ & $99.92\%$ & $99.54\%$ \\ \hline
Token blocking (dirty)&$98.38\%$&$89.99\%$&$94.85\%$&$87.95\%$&$77.34\%$ & $99.92\%$ & $99.54\%$ \\ \hline
Attribute clustering blocking&$97.31\%$&$68.42\%$&$92.10\%$&$76.84\%$&$71.11\%$ & $99.55\%$ & $99.54\%$\\ \hline
Prefix-Infix(-Suffix) blocking (clean-clean)&\multirow{2}{*}{$100\%$}&\multirow{2}{*}{$91.71\%$}&\multirow{2}{*}{$87.68\%$}&\multirow{2}{*}{$95.44\%$}&\multirow{2}{*}{$68.17\%$} & \multirow{2}{*}{N/A} & \multirow{2}{*}{$99.54\%$} \\ \hline 
Prefix-Infix(-Suffix) blocking (dirty)&$100\%$&$89.25\%$&$87.06\%$&$95.50\%$&$74.12\%$ & N/A & $99.54\%$ \\ \hline 

\multicolumn{8}{l}{Precision:} \\ \hline
Token blocking (clean-clean)&$1.56 \cdot 10^{-6}$&$1.00 \cdot 10^{-6}$&$2.49 \cdot 10^{-6}$&$1.30 \cdot 10^{-6}$&$1.13 \cdot 10^{-6}$&$1.21  \cdot 10^{-4}$ & $1.18\cdot 10^{-6}$\\ \hline

Token blocking (dirty)&$3.64 \cdot 10^{-7}$&$5.14 \cdot 10^{-9}$&$3.78 \cdot 10^{-7}$&$1.05 \cdot 10^{-8}$&$1.29 \cdot 10^{-9}$&$7.51  \cdot 10^{-5}$ & $6.5 \cdot 10^{-9}$ \\ \hline

Attribute clustering blocking&$8.51 \cdot 10^{-6}$&$5.76 \cdot 10^{-6}$&$1.01 \cdot 10^{-5}$&$1.41 \cdot 10^{-5}$&$1.35 \cdot 10^{-6}$&$1.52  \cdot 10^{-4}$ & $1.97 \cdot 10^{-6}$\\ \hline

Prefix-Infix(-Suffix) blocking (clean-clean)&\multirow{2}{*}{$1.87 \cdot 10^{-6}$}&\multirow{2}{*}{$2.19 \cdot 10^{-6}$}&\multirow{2}{*}{$5.72 \cdot 10^{-6}$}&\multirow{2}{*}{$1.01 \cdot 10^{-5}$}&\multirow{2}{*}{$2.05 \cdot 10^{-6}$}& \multirow{2}{*}{N/A} & \multirow{2}{*}{$1.19 \cdot 10^{-6}$}\\ \hline 

Prefix-Infix(-Suffix) blocking (dirty)&$6.04 \cdot 10^{-7}$&$8.21 \cdot 10^{-9}$&$2.77 \cdot 10^{-7}$&$1.23 \cdot 10^{-8}$&$6.99 \cdot 10^{-10}$& N/A & $6.84 \cdot 10^{-9}$ \\ \hline 

\multicolumn{8}{l}{F-measure:} \\ \hline
Token blocking (clean-clean)&$3.13 \cdot 10^{-6}$&$2.00 \cdot 10^{-6}$&$9.72 \cdot 10^{-7}$&$2.06 \cdot 10^{-8}$&$1.94 \cdot 10^{-9}$&$2.42 \cdot 10^{-4}$ & $2.35 \cdot 10^{-6}$ \\ \hline
Token blocking (dirty)&$7.28 \cdot 10^{-7}$&$1.03 \cdot 10^{-8}$&$7.55 \cdot 10^{-7}$&$2.10 \cdot 10^{-8}$&$2.59 \cdot 10^{-9}$&$1.50 \cdot 10^{-4}$ & $1.30 \cdot 10^{-8}$ \\ \hline
Attribute clustering blocking&$1.70 \cdot 10^{-5}$&$1.15 \cdot 10^{-5}$&$2.02 \cdot 10^{-5}$&$2.82 \cdot 10^{-5}$&$2.69 \cdot 10^{-6}$&$3.04 \cdot 10^{-4}$ & $3.94 \cdot 10^{-6}$ \\ \hline
Prefix-Infix(-Suffix) blocking (clean-clean)&\multirow{2}{*}{$3.75 \cdot 10^{-6}$} &\multirow{2}{*}{$4.38 \cdot 10^{-6}$}&\multirow{2}{*}{$9.98 \cdot 10^{-7}$}&\multirow{2}{*}{$2.02 \cdot 10^{-5}$}&\multirow{2}{*}{$4.11 \cdot 10^{-6}$} & \multirow{2}{*}{N/A} & \multirow{2}{*}{$2.38 \cdot 10^{-6}$} \\ \hline 
Prefix-Infix(-Suffix) blocking (dirty)&$1.21 \cdot 10^{-6}$&$1.64 \cdot 10^{-8}$&$5.55 \cdot 10^{-7}$& $2.46 \cdot 10^{-8}$ & $1.40 \cdot 10^{-9}$ & N/A & $1.37 \cdot 10^{-8}$ \\ \hline 

\multicolumn{8}{l}{$H3R$:} \\ \hline
Token blocking (clean-clean)&$93.18\%$&$89.55\%$&$95.77\%$&$88.61\%$&$80.90\%$&$70.53\%$ & $97.04\%$\\ \hline
Token blocking (dirty)&$94.05\%$&$90.43\%$&$93.75\%$&$88.97\%$&$86.25\%$&N/A ($RR < 0$) & $90.48\%$ \\ \hline
Attribute clustering blocking&$97.55\%$&$80.76\%$&$95.37\%$&$86.66\%$&$80.80\%$&$79.95\%$ & $98.16\%$ \\ \hline
Prefix-Infix(-Suffix) blocking (clean-clean)&\multirow{2}{*}{$96.09\%$}&$92.70\%$&\multirow{2}{*}{$92.70\%$}&\multirow{2}{*}{$97.18\%$}&\multirow{2}{*}{$79.83\%$}& \multirow{2}{*}{N/A} & \multirow{2}{*}{$97.07\%$}\\ \hline 
Prefix-Infix(-Suffix) blocking (dirty)&$95.92\%$&$90.33\%$&$88.91\%$&$93.59\%$&$83.50\%$& N/A & $90.98\%$ \\ \hline
%
%

\end{tabular}
\label{tab:recall}
\end{table*}

%% file: related_work.tex
\section{Related Work}
\label{sec:related_work}
In this section, we briefly overview representative ER techniques that have been proposed in the Database and Semantic Web communities and can be classified under two general axes: (a) \emph{high} and \emph{low similarity} in the \emph{content} and (b) \emph{high} and \emph{low similarity} in the \emph{structure} of entity descriptions. 

More precisely, \emph{deduplication} techniques~\cite{DBLP:journals/tkde/Christen12} are essentially ER techniques for highly (low) similar in structure (content) descriptions from one relation, \emph{record  linkage} for structured~\cite{DBLP:journals/tkde/Christen12} or semi-structured Web data~\cite{hassanzadeh2013record} targets highly (low) similar in content (structure) descriptions from two relations, while in the Web of data, descriptions hosted in a network of KBs exhibit low similarity both in content and structure (i.e., are \emph{somehow} similar). It is worth noticing that \emph{instance matching} techniques~\cite{nentwig2015survey}  
(also called data linking or link discovery), aiming to semantically relate entities described in two KBs, are variations of the record  linkage problem for semi-structured descriptions (e.g., in RDF) and relationships going beyond entity equivalence (e.g., sameAs). Web-scale ER techniques essentially require to address these problems for a large number of entity types and hosting KBs, having limited domain knowledge regarding how description schemas and instances could match, as well as representative ground truth and training sets. 
In this context, resolution decisions regarding one pair of descriptions essentially provide evidence for  the matching of others (i.e., the network effect).

In the rest of this section, we focus on blocking (and meta-blocking) methods outside the scope of our work and explain the reasons why they have not been included in our experimental evaluation. We finally compare our work with other benchmarking and large-scale evaluation efforts.


\textbf{Blocking. } 
Besides the methods we detailed in Section~\ref{sec:preliminaries}, alternative methods for reducing the number of unnecessary comparisons include: Locality-Sensitive Hashing, similarity joins, frequent itemsets, and clustering.

The key idea of blocking with Locality-Sensitive Hashing (LSH) (e.g.,~\cite{DBLP:conf/edbt/MalhotraAS14}) is to hash descriptions multiple times, using a family of hash functions, in such a way that similar descriptions (e.g., with Jaccard similarity, approximated by  minhasing~\cite{DBLP:journals/jcss/BroderCFM00}) are more likely (with probabilistic guarantees) to be placed into the same bucket than dissimilar ones. Any two descriptions that hash at least once into the same bucket, for any of the employed hash functions, are considered to be a candidate pair. This technique assumes an a-priori knowledge of a minimum similarity threshold between entity description pairs, above which, such pairs are considered candidate matches. However, as we will see in our experimental evaluation (see Section~\ref{sec:experiments}), often, matching descriptions do not share many common tokens and thus, have very low, even zero, similarity when computed only on the values of their attributes. 
Those matches would not be placed in the same bucket and thus, they would not be considered candidate matches. 
Effectively choosing a minimum similarity threshold also depends on the KBs. For example, when seeking matches between two central KBs, a high similarity threshold can be used, since such KBs usually have more similar values. Using a lower threshold in central KBs would result in many false candidate pairs. Accordingly, using a high similarity threshold in peripheral KBs, in which descriptions have lower similarity values, would yield many missed matches. 
Consequently, applying LSH across domains is an open research problem, due to the difficulty in knowing or tuning a similarity threshold that can be generalized to identify matches across several domains in an effective and efficient way. 


String-similarity join algorithms (e.g.,~\cite{DBLP:conf/icde/ChaudhuriGK06,DBLP:conf/www/BayardoMS07,DBLP:conf/www/XiaoWLY08}) construct blocks which are guaranteed to contain all pairs of descriptions whose string values similarities are above a certain threshold and potentially some pairs whose string values similarities are below that threshold. To achieve that, without computing the similarity of all pairs of descriptions, this family of algorithms build an inverted index from the tokens of the attribute values of the descriptions. However, unlike token blocking, this inverted index is created only by the first non-frequent tokens of each description (i.e., the most discriminating), based on the \emph{prefix filtering} principle~\cite{DBLP:conf/icde/ChaudhuriGK06}. 
\cite{DBLP:conf/www/BayardoMS07} additionally applies a \emph{size filtering}~\cite{DBLP:conf/vldb/ArasuGK06} on the sets of tokens to disregard some of the candidate pairs, based on the fact that $Jaccard(x,y) \geq t \Rightarrow t \cdot |x| \leq |y|$.
The ppjoin+ algorithm~\cite{DBLP:conf/www/XiaoWLY08} introduces a \emph{positional filtering}, i.e., the position in the ordered set of tokens, in which a token appears, to further reduce the number of candidate pairs. 
Tuning the appropriate similarity threshold is non-trivial and it also affects the performance of the string-similarity join algorithms~\cite{DBLP:journals/pvldb/JiangLFL14}. Smaller thresholds entail less pruning, and thus, more time. Furthermore, \cite{DBLP:journals/pvldb/MetwallyF12}  proves experimentally that algorithms based on prefix filtering are only effective when the similarity threshold is extremely high. However, this is not the case in the Web of data, where highly heterogeneous descriptions, yielding very low similarity in their literal values, can refer to the same entity. 



\cite{DBLP:journals/is/KenigG13} introduces a method for building blocks based on Maximal Frequent Itemsets (MFI). Abstractly, each MFI (an itemset can be a set of tokens) of a specific attribute in the schema of a description defines a block, and descriptions containing the tokens of an MFI for this attribute are placed in a common block. Using frequent itemsets to construct blocks may significantly reduce the number of candidates for matching pairs. 
However, since many matching descriptions share few, or even no common tokens, further requiring that those tokens are parts of frequent itemsets is too restrictive for those pairs of matching descriptions, resulting in many missed matches in the Web of data. Moreover, MFI blocking requires a-priori knowledge of the desired block sizes, and is also based on the notion of a schema, information which is unavailable at the Web of data. 


Canopy clustering~\cite{DBLP:conf/kdd/McCallumNU00} is an unsupervised clustering method that has been used as blocking. Initially, one random description is chosen as a canopy center and it is compared to all other descriptions. All the descriptions within a loose distance threshold to this canopy center are added to this canopy, while those within a tighter distance threshold, are removed from the candidate canopy centers. This process repeats, until there are no more candidate canopy centers. 
The problem with this approach, even with the parallel version of  Mahout\footnote{\url{mahout.apache.org/users/clustering/canopy-clustering}} or our own MapReduce implementation, is that it fails to scale to the volumes of our datasets.

\textbf{Block post-processing.} 
Further processing steps on the results of blocking have been proposed in the literature for further reducing the number of comparisons to be performed by an ER task (e.g.,~\cite{DBLP:journals/tkde/PapadakisKPN14,DBLP:conf/bigdataconf/Efthymiou0PSP15,DBLP:journals/tkde/WhangMG13}). Such steps make sense to be used, when blocking results in missing only few matches, and the whole process is faster than exhaustively performing the comparisons between all descriptions.  
For example, \cite{DBLP:journals/tkde/PapadakisKPN14} proposes to reconstruct the blocks of a given blocking collection in order to discard redundant, as well as unnecessary comparisons.  \cite{DBLP:journals/tkde/PapadakisKPN14} essentially transforms a given blocking collection \hide{$B$} into a blocking graph\hide{$G_B$}, whose nodes \hide{$V_B$} correspond to entity descriptions\hide{in $B$}, while its undirected edges \hide{$E_B$} connect the co-occurring descriptions. 
Every edge \hide{$e_{i,j}$} is associated with a weight \hide{$w_{i,j} \in [0,1]$} representing the likelihood that the adjacent entities are matching candidates. Low-weighted edges are pruned, so as to discard comparisons between unlikely to match descriptions. 
To minimize the missed matches, an iterative entity resolution process can exploit in a {\em pay-as-you-go} fashion any intermediate results of blocking and matching, discovering new candidate matches. Such iterative process may consider matching evidence provided by entity descriptions placed into the same block (e.g.,~\cite{DBLP:conf/sigmod/WhangMKTG09}) or being structurally related in the original entity graph (e.g.,~\cite{DBLP:conf/kdd/Lacoste-JulienPDKGG13,DBLP:journals/pvldb/RastogiDG11}). 
In this work, we are focusing on blocking algorithms, since those further steps also benefit from a better initial blocking, but we briefly show how such an algorithm ~\cite{DBLP:conf/sigmod/WhangMKTG09} uses the results of blocking.

\textbf{Benchmarks and evaluation frameworks.} 
Existing works in ER benchmarks~\cite{DBLP:journals/jodsn/IoannouRV13,DBLP:conf/esws/FerraraMNS11} and evaluation frameworks \cite{DBLP:journals/pvldb/KopckeTR10,DBLP:journals/pvldb/HassanzadehCML09} focus on the similarity of descriptions and how these similarities affect the matching decision of entity resolution; not on blocking, explicitly. In all cases, data collections are built from central datasets of a single domain, e.g., only bibliographic. Those data variations are not adequate to evaluate the blocking algorithms suitable for cross-domain ER involving a large number of entity types. 
Finally, many works on ontology and instance matching, e.g., \cite{DBLP:journals/pvldb/SuchanekAS11,DBLP:conf/kdd/Lacoste-JulienPDKGG13}, have been using the OAEI benchmarks in their evaluations. 
Typically, those collections are composed of two ontologies with a 1-1 mapping in their attributes, or even a single ontology, whose instances, i.e., entity descriptions, have some modifications in their values. 
We have included and analyzed one of those benchmarks in this study.

%% file: summary.tex
\section{Summary}\label{sec:summary}
In this work, we evaluated, for the first time, ER blocking algorithms for
somehow (not only highly) similar descriptions in the Web of data.
We have investigated the data characteristics of such descriptions that impact blocking algorithms' effectiveness and efficiency.
Highly similar descriptions, met in central LOD collections, feature
many common tokens in the values of common attributes, while somehow
similar descriptions, met in peripheral collections, have significantly
fewer common tokens in attributes that are not necessarily semantically
related. Hence, the former can be compared only on their content (i.e.,
values), while the latter require contextual information, e.g., the similarity of neighborhood descriptions, linked with different types of relationships.
Since a single similarity function cannot identify such matches in a single
pass, multiple iterations of matching (focusing on context) and/or
blocking (focusing on content) are needed. 
Towards this end, we are interested in progressive ER algorithms that try to maximize the benefit (e.g., number of resolved entities, number of links between resolved entities) of each iteration, by dynamically adapting their execution plan, based on previous results.